\documentclass[letter,9pt]{article}
\usepackage[scale=.83]{geometry}
\usepackage[affil-it]{authblk}
\usepackage{amssymb,amsmath,amsthm,amsbsy}
\usepackage{physics}
\usepackage{graphicx}
\usepackage{epstopdf}
\usepackage{epsfig}
\usepackage{multicol}
\usepackage{caption}
\usepackage{multirow}
\usepackage{algorithm,algorithmic}

\usepackage{setspace}
\usepackage{color}
\usepackage{lineno}
\usepackage{svg}
\usepackage{pdfpages}
\usepackage{subfig}
\usepackage{comment}
\usepackage{cleveref}
\usepackage{booktabs}
\usepackage{ifthen}
\usepackage{microtype} 

\usepackage{titlesec} 
\usepackage[toc]{appendix}
\titleformat{\section}[block]{\large\scshape\centering}{\thesection.}{1em}{} 
\titleformat{\subsection}[block]{\large}{\thesubsection.}{1em}{} 
\usepackage[export]{adjustbox}

\title{\textbf{Higher-order transmissibility and its linear approximation for in-service crack identification in train wheelset axles}}

\author{Ehsan Naghizadeh\footnote{Corresponding author.\\ \normalfont{E-mail address: enaghizadeh@ethz.ch}} $^{,1}$}
\author{Eleni Chatzi$^2$}
\author{Paolo Tiso$^1$}

\affil{
$^1$Institute for Mechanical Systems, ETH Zürich, Switzerland \\$^2$Department of Civil, Environmental and Geomatic Engineering, ETH Zürich, Switzerland }

\date{}

\renewcommand{\vec}[1]{\boldsymbol{\mathrm{#1}}}     
\newcommand{\mat}[1]{\boldsymbol{#1}}           

\renewcommand{\norm}[1]{\| #1 \|}                 

\begin{document}
	

\maketitle 
\thispagestyle{empty}

\noindent\textbf{Abstract}: In-service structural health monitoring is a so far rarely exploited, yet potent option for early-stage crack detection and identification in train wheelset axles. This procedure is non-trivial to enforce on the basis of a purely data-driven approach and typically requires the adoption of numerical, e.g. finite element-based simulation schemes of the dynamic behavior of these axles. Damage in this particular case can be formulated as a breathing crack problem, which further complicates simulation by introducing response-dependent nonlinearities into the picture. In this study, first, a new crack detection feature based on higher-order harmonics of the breathing crack is proposed, termed Higher-Order Transmissibility (HOTr). Next, a reduced-order model is developed by introducing a first-order perturbation-based approximation of the essentially nonlinear crack contact forces. This yields a sequence of linear frequency-domain problems that approximate the higher-order harmonic response with substantial computational speedup, while retaining the damage-sensitive nonlinear features required for crack identification. The accuracy of the proposed method in reproducing the delivered HOTr is compared against the nonlinear simulation model. The obtained results suggest that the approximation of the HOTr can significantly reduce the computational burden by eliminating the need for an iterative solution of the governing nonlinear equation of motion while maintaining a high level of accuracy when compared to the reference model. Finally, the robustness and effectiveness of the proposed indicators are systematically demonstrated through noise-contaminated simulations. Results advocate the great potential of the proposed method for adoption in in-service damage identification for wheelset axles, feasibly within a near real-time context.


\section{Introduction}

The structural integrity of the wheelsets comprises an important aspect of the condition monitoring of trains. The high risk of train derailment as a result of a failure in the wheelset advocates the development of efficient schemes for Structural Health Monitoring (SHM) of the wheelset axles\cite{madler1,pp}. Incipient cracks, which cause high cycle fatigue failure due to crack propagation, form a main threat to the integrity of train wheelsets \cite{cantini2011structural,zerbst2013safe}. As a result, \emph{early-stage} crack detection and identification in train wheelset axles have served as the subject of investigation for numerous research studies in the past century \cite{Sabnavis2004,sekhar1998condition}.

The most common industry practice for crack identification lies in the adoption of Non-Destructive Testing (NDT) schemes, which are usually applied within a dedicated workshop facility. In particular, Ultrasound Testing (UT) is shown to be capable of detecting the so-called incipient cracks, which are sized between 1-5\% of the shaft diameter \cite{benyon2001use,Zerbst2013}. However, the implementation of NDT methods, including UT, requires the train to be stationed and, thus, out of service for a prolonged inspection period, while further necessitating sophisticated equipment and high-cost facilities. In such a setting, the inspection intervals are prescribed at scheduled periods, while the optimal planning of these intervals, which is likely non-regular, poses a non-trivial task requiring a verifiable residual life estimation of the wheelset \cite{madler1,Shu2024,Yan2023}. Studies on crack propagation and residual life estimation of the wheelset axle under operational conditions form an ongoing research topic that is challenged by various factors such as, among others, the temporal and spectral nature of the applied external load \cite{simunek2018situ,maierhofer2020fatigue}, the involved material characteristics \cite{rieger2020fatigue}, crack closure \cite{leitner2019retardation,maierhofer2018oxide} and the possible development of residual stresses \cite{pertoll2024residual}. The lack of precise knowledge about the mentioned parameters results in associated uncertainties and prompts the adoption of conservative safety factors for scheduling appropriate inspection intervals \cite{madler1}. An alternative approach to the current practice lies in the broader exploitation of in-service health monitoring schemes. Such techniques rely on the continuous evaluation of axle integrity through real-time data from a set of sensors implemented on the axle. The continuous surveillance scheme reduces the risk of axle failure, increases the availability of trains due to reduced maintenance, and, most importantly, eliminates the need for a verifiable crack growth model. However, such an approach is challenged by several factors, such as in-service data acquisition (the difficulty in mounting permanent sensors on the rotating axle, as achieved in \cite{Beretta2016,maglio2022railway,bracciali2014review}), the extraction of robust crack identification features, and the need to run the identification procedure in real or close to real time, thus facilitating data compression and lowering storage and transmission requirements.

Unlike damage detection, crack identification techniques require information from a model of the structure that is parameterized with respect to the crack characteristics (location, size, and orientation) and is also capable of predicting the response of the shaft to varying operational loads \cite{adams2007health}. Identification is then typically achieved by solving an inverse problem through optimization to minimize the error between a Damage Sensitive Feature (DSF) derived from the measured response and the corresponding value calculated using the model. The mentioned models can be largely classified into either the purely data-driven or physics-based classes. The primary obstacle for the former is the actual availability of high-quality data \cite{AVCI2021107077}. Specifically, in the case of axle cracks, such data are not available due to the rarity of such damage occurrences. To alleviate such a hurdle, data-driven models could be trained on synthetic data derived from high-fidelity physics-based models. Nevertheless, seeding a large pool of solutions to span the parameter space of the problem often requires the repetitive evaluation of the high-fidelity model for a prohibitively large number of times, rendering the process inefficient. Physics-based models, on the other hand, do not require training for the evaluation of a forward problem. However, during the solution of the inverse problem, the model must be evaluated a significant number of times. Therefore, a computationally efficient model is of paramount importance.

Substantial literature exists on crack detection and identification within the context of vibration-based monitoring and non-destructive evaluation. The interested reader is referred to the works of Carden \& Fanning \cite{Carden2004} and Sabnavis \textit{et al.} \cite{Sabnavis2004}. Here, we confine our analysis to the availability of vibration-based monitoring data, owing to the merits and applicability of associated sensors, such as accelerometers and strain gauges, for in-service surveillance \cite{Sun2023}. A comprehensive review of traditional vibration-based damage identification is provided in \cite{fan2011vibration}. An issue commonly faced by conventional DSFs related to vibration-based monitoring, such as modal properties, mode shape curvatures, transmissibility functions, and structural flexibility, is their insensitivity to small cracks \cite{sinou2009review}. Ideally, a DSF that is suited for crack detection and characterization should be both global, in the sense that it can be detected from sparse measurements, and simultaneously local to the crack, i.e., exhibiting sufficient sensitivity to the crack characteristics (location and size). Toward extracting such DSFs, crack breathing phenomena have been capitalized upon to extract the so-called higher-order harmonics as features that meet the aforementioned criteria \cite{Hiwarkar2012,lin2018higher}. In this context, Reduced-Order Modeling (ROM) can be used as a framework that supports the affordable simulation of associated crack models, which are often computationally taxing. However, such ROMs often struggle to cover the large parameter space typically encountered in crack identification problems \cite{agathos2024accelerating}.

The implementation of breathing-induced higher-order harmonics for online SHM dates back to 1989 \cite{Imam1989}, and has found applications for crack detection in different structures \cite{Tsyfansky2000}. A thorough review of damage modeling and identification techniques is given in \cite{abdul2024modeling}. Two main approaches exist for simulating breathing cracks. The first is to model the alternating bending stiffness due to the crack as a harmonic function of time that is synchronized with the rotational speed of the axle \cite{pilkey2002analysis,guo2013application}. Although the mentioned methods are computationally less expensive, they provide very limited resolution for crack parameters and are generally used for a qualitative study of cracked axles. The second, more accurate approach is based on the adoption of a Finite Element (FE) model of the axle with a fine mesh and explicit modeling of the contact between the crack faces. The crack can then be represented either through meshless techniques, such as the eXtended Finite Element Method (XFEM) \cite{agathos2018multiple}, or by confining the crack shape to the underlying mesh, coupled with appropriate parameterization schemes such as mesh morphing \cite{Agathos2021Parametric}. The dependency of the crack behavior on the contact state of the crack faces renders the analysis nonlinear. The fundamentals underlying the nonlinear analysis of a shaft with a breathing crack are presented in \cite{Wang2018,el2019nonlinear}. In recent years, Lin \& Ng implemented a higher-order Frequency Response Function (FRF) for the identification of localized nonlinearities in structures, with a focus on breathing cracks \cite{Lin2018}. The effect of imbalance and its interaction with higher-order harmonics stemming from the breathing phenomenon is examined in \cite{Kushwaha2023,Han2024}. This concept was later extended to the identification of multiple cracks \cite{Chomette2020}. The experimental study presented in \cite{guo2017experimental} provides validation for employing higher-order harmonics in crack detection. The findings indicate the presence of higher-order harmonics, extending up to the fifth order. The identification of early-stage cracks in rotating axles can be efficiently achieved through the anti-resonance of higher-order harmonics, as demonstrated in \cite{Sinou2022ant}. Subsequently, in \cite{Sinou2022}, the authors enhance this crack identification approach by incorporating model uncertainties, thereby improving the robustness of the identification process. In an additional numerical study, the investigators demonstrate that up to fourth-order harmonics appear in a rotating hollow shaft with cracks \cite{lu2016nonlinear}. Higher-order harmonics have recently been shown to comprise robust features for detecting breathing cracks in fluid-delivering pipes \cite{Ji2023}. Relying on the breathing phenomena authors introduce a new metric based on the differences of the vibrating frequencies in open and close phase of crack breathing in\cite{wei2023novel}. The proposed method shows stronger robustness against noise than the conventional methods.

To consider the interaction between crack faces, a nonlinear analysis is required, which adds to the computational complexity involved in evaluating a forward model. The need for fast calculation of the model response, especially within inverse problem settings, remains a major challenge and the main repelling factor for implementing the nonlinear DSF for the in-service SHM of the wheelset axles. In this work, we introduce the novel concept of Higher-Order Transmissibility (HOTr) as a crack detection feature to address the problems associated with the high computational cost of the aforementioned nonlinear analysis. HOTr is derived by extending the concept of the transmissibility function that has been advocated in the literature for open cracks (i.e., without contact) \cite{farrar2012structural,chesne2013damage,yan2019transmissibility} to higher-order harmonics induced by breathing cracks, as described in Section \ref{section: Methods}. We show that the crack face contact force can be approximated by the constraint forces of a closed crack, resorting to linear system theory. The latter results in a significant reduction in the required computational time for obtaining a sufficient approximate solution. The efficiency of the proposed DSF for crack identification is tested on two numerical examples.

The remainder of the paper is organized as follows. In section \ref{sec:postulate}, we postulate that the presence of a breathing crack leads to higher-order harmonics in the response signal. In section \ref{section: Methods} detailed mathematical modeling of the FE discretized axles with breathing cracks is presented, the HOTr is defined, their linear approximation is formulated, and the inverse problem for crack identification based on the HOTr is presented. Section \ref{section: numerical examples} discusses two numerical examples by first evaluating the sensitivity of the proposed HOTr to crack parameters, then inspecting the effectiveness of the proposed method in approximating HOTrs; and finally performing crack identification and providing a statistical view of the identified crack parameters. Conclusions on the obtained results are summarized in section \ref{section: conclusion}.

\section{Postulation - Single Degree of Freedom system} \label{sec:postulate}

\begin{figure}[h]
    \centering
    \adjincludegraphics[width=3.5in,trim={0 {0.4\height} {0.6\width} {0.01\height} },clip]{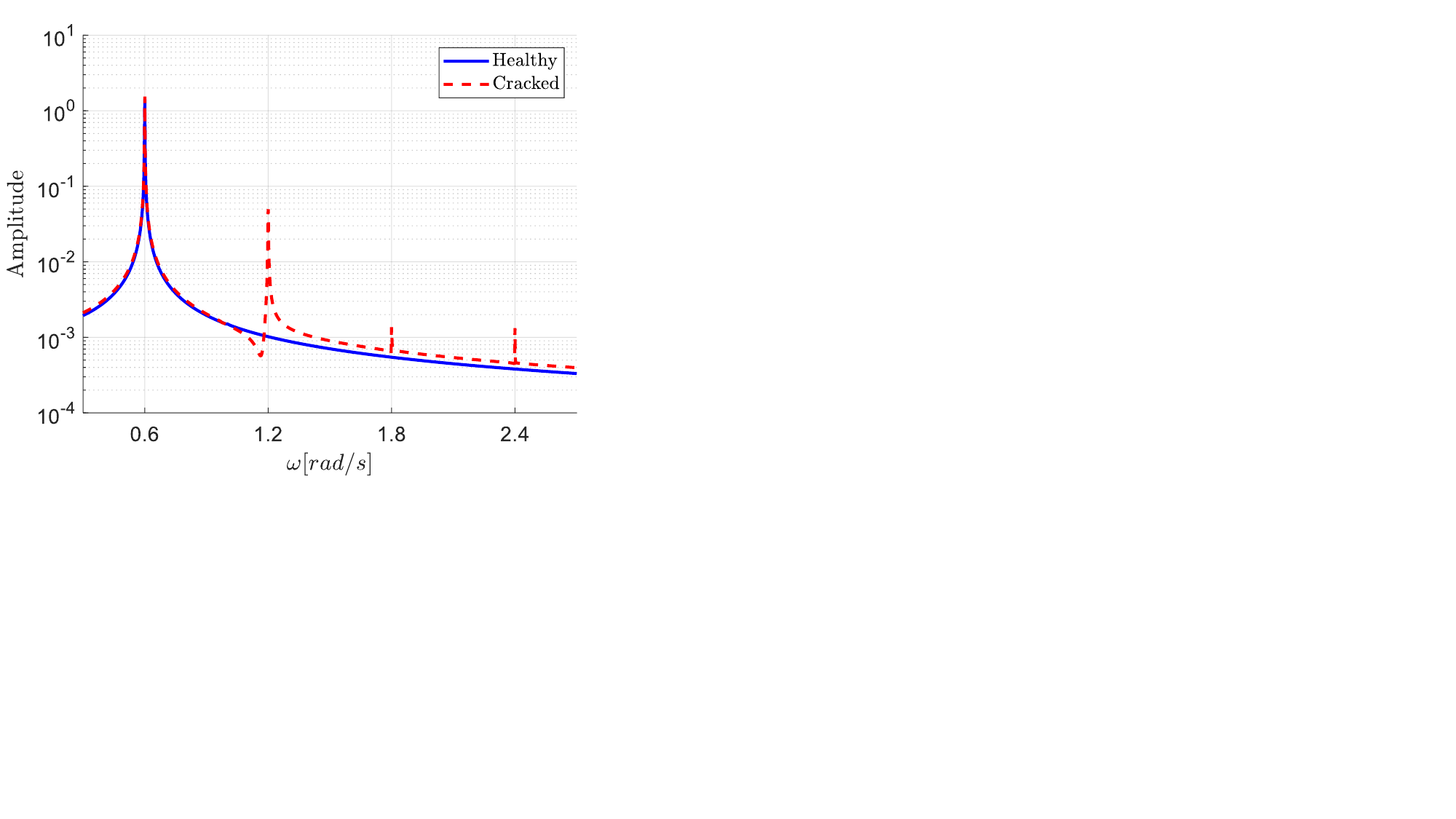}
    \caption{Response spectrum of the bilinear SDoF system representing a cracked axle to harmonic excitation at 0.6 [rad/s] and comparison versus the pristine case. The appearance of higher-order harmonics is notable.}
    \label{fig: SDOF_FRF}
\end{figure}

To illustrate the appearance of higher-order harmonics due to the presence of a breathing crack, let us consider a Single Degree of Freedom (SDoF) mass-spring-damper system featuring an additional unilateral spring subjected to sinusoidal forcing, as shown in Figure \ref{fig: SDOF}. This model qualitatively reproduces the differentiation in bending stiffness due to the opening and closing of the crack as a cracked beam vibrates in its first bending mode. The governing equation of motion can be written as follows:
\begin{figure}[h]
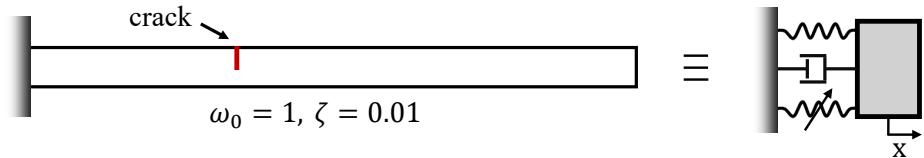

    \centering
    \adjincludegraphics[width=5in,trim={0 {0.8\height} {0.38\width} 0 },clip]{SDOF_model.pdf}
    \caption{SDoF contact model representing bending beam with crack.}
    \label{fig: SDOF}
\end{figure}

\begin{equation}
     \ddot{\text{x}}(\text{t}) + 2\zeta\omega_{0}\dot{\text{x}}(\text{t}) + \omega_{0}^{2}\text{x}(\text{t}) - \epsilon H(\text{x})\omega_{0}^{2}\text{x}(\text{t}) = \text{cos}(\omega \text{t}).
     \label{eq: EOM_SDOF}
\end{equation}
where $\text{x}(\text{t})$ represents the displacement of the beam tip, and $H\text{(\text{x})}$ is the step function defined as:

\begin{equation}
        H\text{(\text{x})} = 
        \begin{cases}            1,& \text{x}\geq 0\\
            0,& \text{x} < 0
            \end{cases}.
    \label{eq: step function}
\end{equation}
The system parameters are chosen  as $\omega_{0} = 1$, $\zeta = 0.01$. The extent of the crack is denoted by the parameter $\epsilon$. We then simulated the response of both pristine, i.e., $\epsilon= 0$ and the case with a small crack, i.e., $\epsilon= 0.01$ using this simple model under harmonic excitation at $\omega_\text{f} = 0.6~[\text{rad/s}]$ with a unit magnitude, as shown in Eq. \ref{eq: EOM_SDOF}. The simulation ran until the steady-state was reached. The resulting spectrum of the response, depicted in Figure \ref{fig: SDOF_FRF}, reveals the presence of higher-order harmonics that can be traced to the breathing phenomenon. Moreover, it is observable that the first-order harmonic does not exhibit any alterations, demonstrating its insensitivity to incipient cracks. The higher-order harmonics are distinctive features attributed to the breathing phenomena that arise from the crack, making these a suitable DSF for crack identification. The extraction of the mentioned harmonics from noisy acceleration and/or strain measurements presents further challenges due to the low amplitude associated with these features, as seen in Figure \ref{fig: SDOF_FRF}. In this work, we focus on the second-order harmonic, which is expected to have more notable energy than higher orders (e.g., 3,4, etc.).

\section{Methods}
\label{section: Methods}
\subsection{Train axle with breathing crack in operation - Equation of Motion}

\begin{figure}[h]
    \centering
    \adjincludegraphics[width=5cm,trim={0 {0.72\height} {0.79\width} 0 },clip]{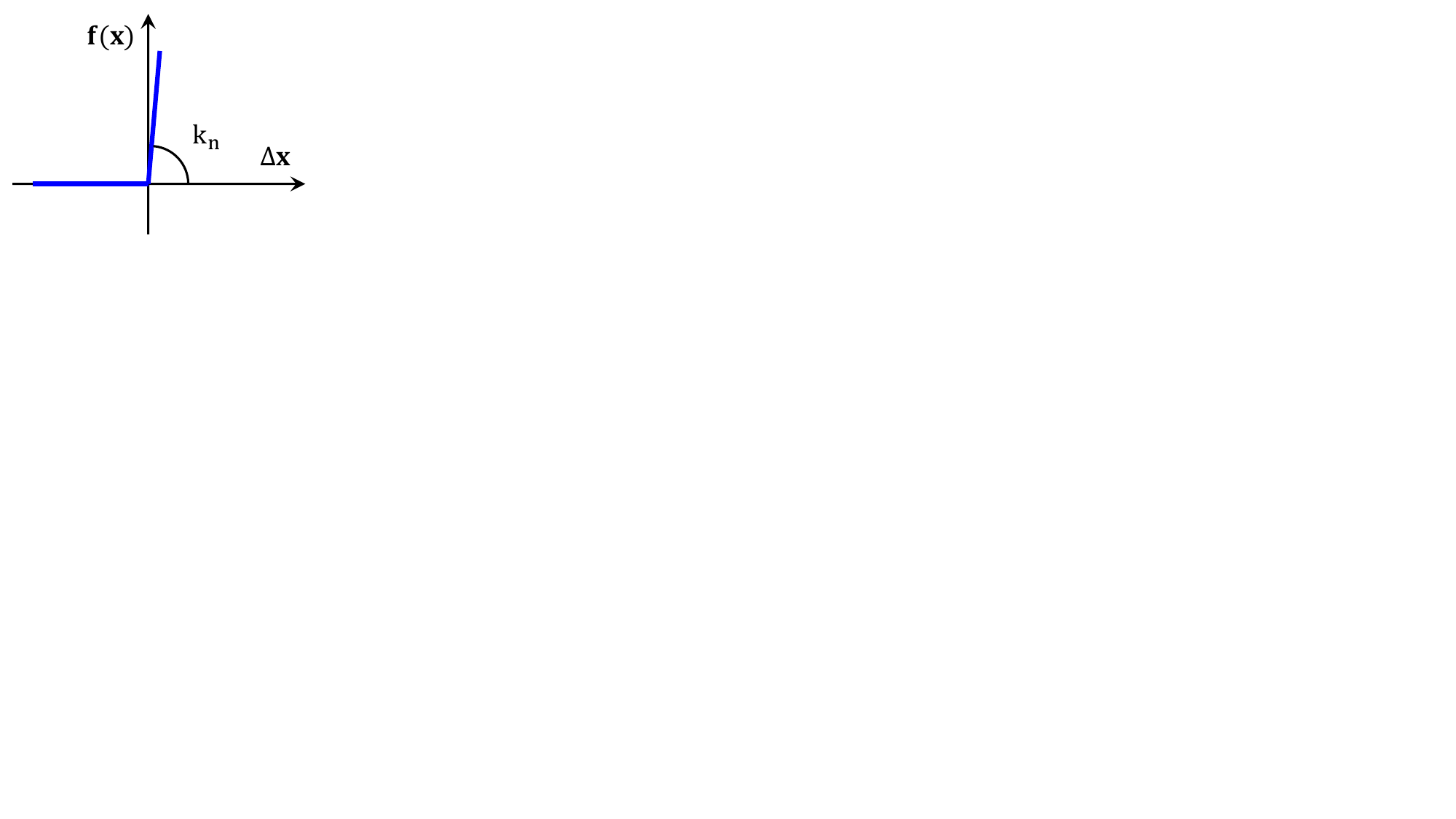}
    \caption{Schematics of the penalty contact force against the relative displacement at the crack face.}
    \label{fig: contact_force}
\end{figure}

\begin{figure}[h]
    \centering
    \adjincludegraphics[width=5.5in,trim={0 {0.4\height} {0.32\width} 0 },clip]{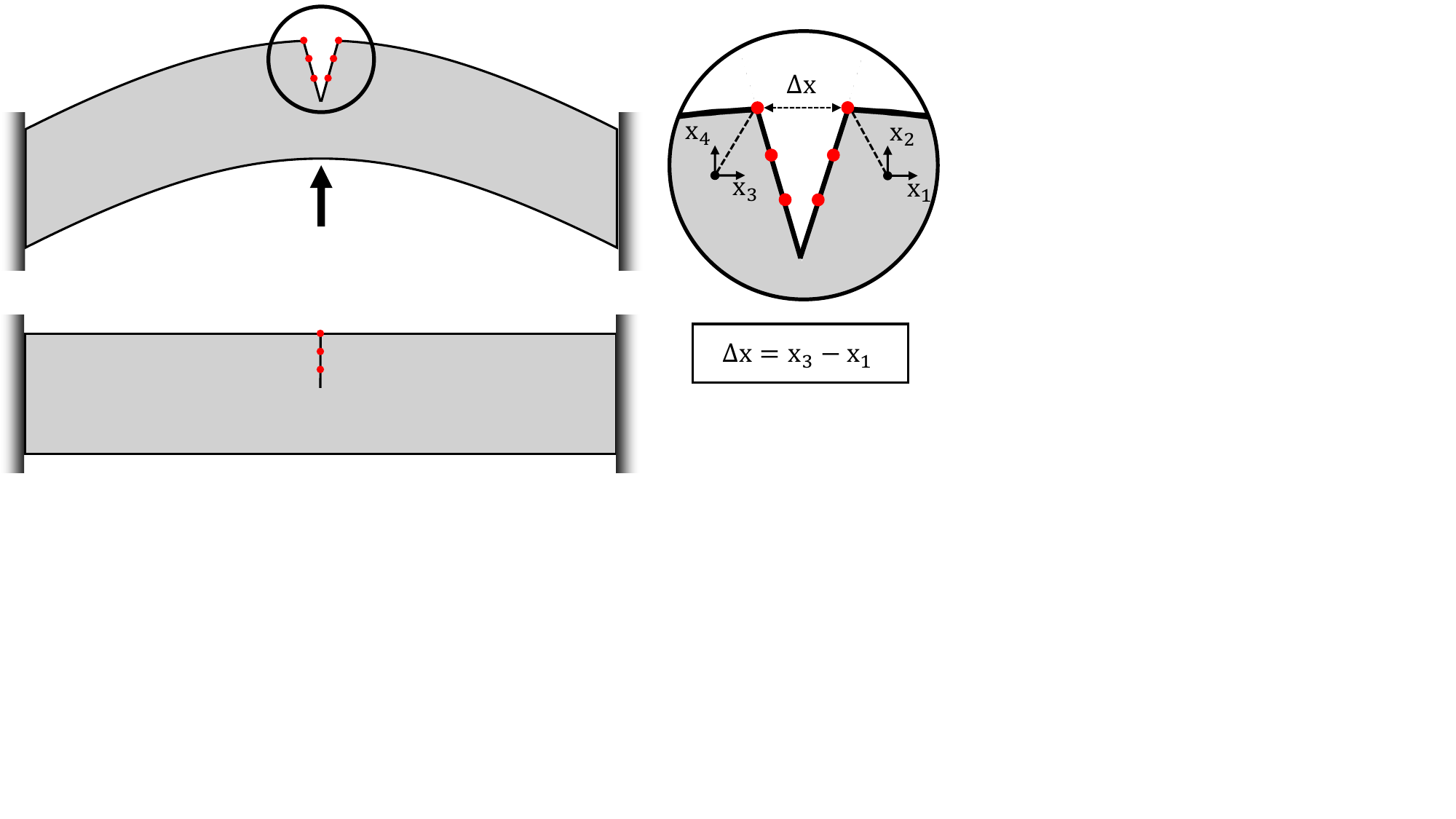}
    \caption{Normal relative displacement ($\Delta\text{x}$) for a contact pair}
    \label{fig: open_crack}
\end{figure}

Different approaches exist for modeling a wheelset axle, depending on the goal of the analysis. In this study, we focus on the nonlinearity induced by the breathing crack as a damage identifier. Therefore, the devised model should be crafted in a manner that intentionally captures the nonlinear features associated with the breathing phenomenon. To this end, a finite element model of a train axle is adopted. The main reason for this choice is that such a discretized numerical approximation allows for explicit parameterization of the crack with respect to its location and size. Initially, the finite element model of the \emph{open} crack is constructed with a conforming mesh on the opposing crack faces. Subsequently, the breathing mechanism is implemented by applying contact between the crack surfaces.(see Figure \ref{fig: open_crack}). The proposed method is a well-established modeling strategy for breathing cracks and has been extensively adopted and validated in the literature for simulating crack opening–closing behavior in rotating shafts and beams~\cite{oh2020stability,wang2018effects,bachschmid2010cracked,georgantzinos2008insight} . The equation of motion can then be written as:

\begin{equation}
     \mat{M}\ddot{\vec{x}}\text{(t)} + \mat{C}\dot{\vec{x}}\text{(t)} + \mat{K}\vec{x}\text{(t)} + \hspace{0.02cm}\vec{f}(\vec{x}\text{(t)}) = \vec{q}\text{(t)}.
     \label{eq: EOM}
\end{equation}
In Eq. \ref{eq: EOM}, $\vec{x(\text{t})}\in\mathbb{R}^{r+2c}$ is the vector of displacements at the discretized degrees of freedom (DoFs), where $c$ is the number of DoFs at each crack face and $r$ is the number of all other DoFs (see Figure \ref{fig: open_crack}). The mass, damping, and stiffness matrices are denoted as $\mat{M}$, $\mat{C}$, and $\mat{K}$ respectively. The contact between the crack faces is modeled using the penalty method by introducing a node-to-node piecewise linear contact law. The resulting bilinear force, oriented normal to the crack face, and exerted between a pair of nodes at the crack faces (see Figure \ref{fig: open_crack}), $\vec{f}(\vec{x}\text{(t)})$ can be formulated as

\begin{equation}
        \vec{f}(\vec{x}\text{(t)}) = -\text{k}_\text{n}\mat{B}^{T}H(\Delta \vec{x}) \circ \Delta \vec{x}, \hspace{0.5cm} \text{where} \hspace{0.15cm}\vec{f}_{j} = \text{k}_\text{n}\mat{B}_{j}^{T}H(\Delta \vec{x}_j) \Delta \vec{x}_j.
        \label{eq:contact_law}
\end{equation}
In Eq. \ref{eq:contact_law}, $\text{k}_\text{n}$ is a scalar penalty value, $\Delta \text{x} \in\mathbb{R}^{p}$ is the vector of relative normal displacements between the  pairs of contact nodes (see Figure \ref{fig: open_crack}) with $p$ being the total number of contact pairs. Finally, $\mat{B}\in \mathbb{R}^{(r+2c)\times p}$ is the signed boolean matrix such that $\mat{B}\vec{x} = \Delta\vec{x}$. A schematic of the contact force, $\vec{f}(\vec{x})$, is shown in Figure \ref{fig: contact_force}.

Additional damping effects arising from friction or local contact hysteresis are assumed to be negligible and are therefore omitted. Throughout the study, it is assumed that the nonlinearity is confined to the nonlinear forces at the crack faces, and there exist no other sources of nonlinearity in the system.

\subsection{Steady-state response and higher-order harmonics}

\begin{figure}[h]
    \centering
    \adjincludegraphics[width=2.5in,trim={0 {0.65\height} {0.74\width} 0 },clip]{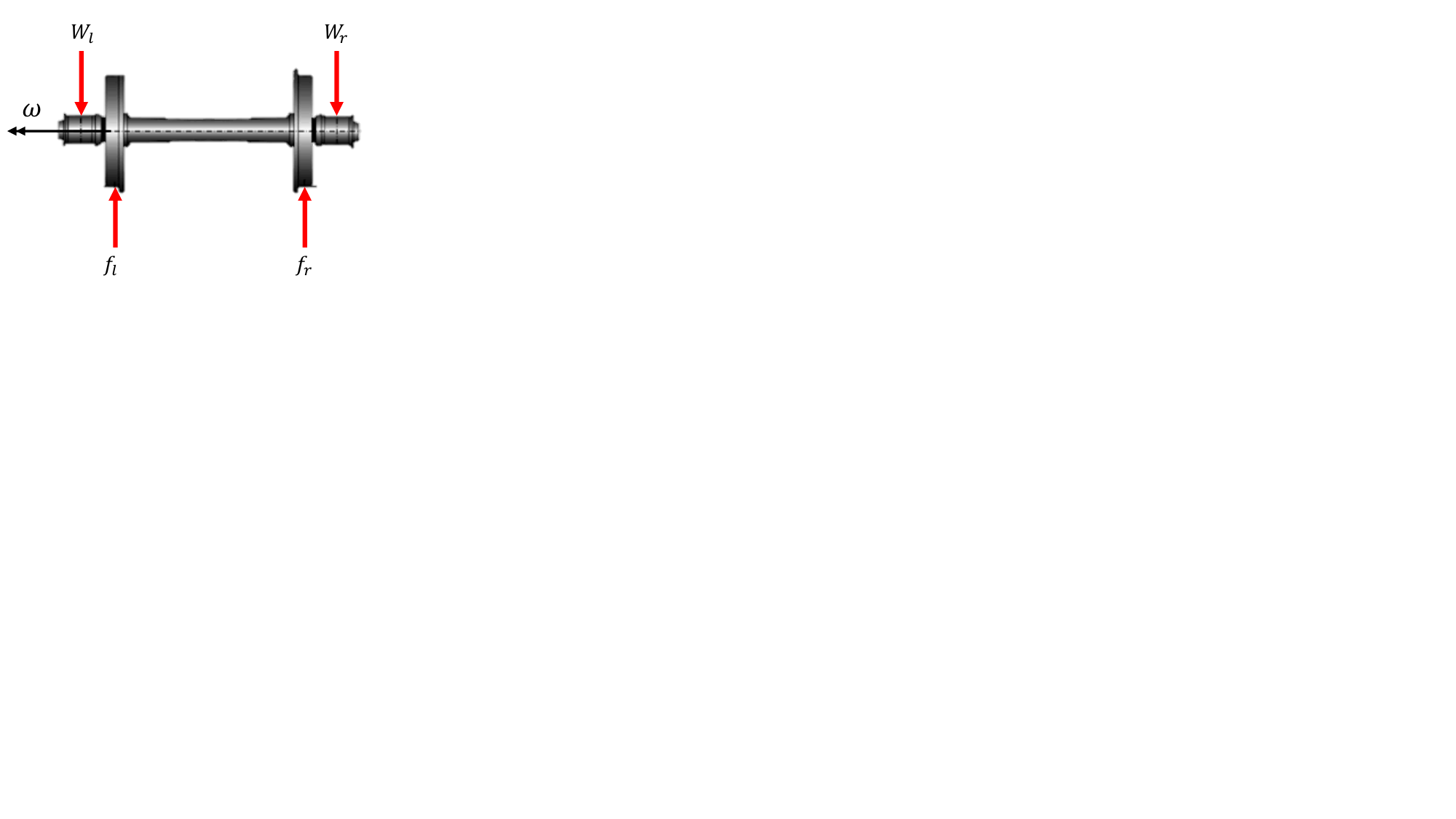}
    \caption{Free body diagram of a wheelset axle in operation. $W_l$ , and $W_r$ denote the load from the weight of the wagon coach. The reaction force form the interaction of the wheels with the rail is denoted as $f_l$ , and $f_r$.}
    \label{fig: loads}
\end{figure}

The axle is assumed to be subject to a four-point bending load due to the weight of the train coach and the reaction force from the rails, as shown in Figure \ref{fig: loads}. The weight distribution (i.e. $W_l$, and $W_r$) is assumed to be known and time-invariant for the time window of interest. An approximation of this distribution can be obtained from strain measurements on the axle. In normal operating conditions, the wheelset rolls without slip, with an angular velocity $\omega$ proportional to the train speed. 

The equation of motion can then be written with respect to a rotating reference frame attached to the axle and rotating with it. In this reference frame, the bending load rotates with $\omega$. By considering a constant speed for the train, the external forces exerted on the axle become a single-tone harmonic excitation with frequency $\omega$. Note that additional stresses due to the rotation of the axle, i.e., rotordynamic effects and the whirl motion caused by the eccentricity of the axle, are neglected here due to the high rigidity and low operating rotational speed of the train axles. Then, the external force $\vec{q}\text{(t)}$ can be expressed as

\begin{equation}
    \vec{q}\text{(t)} = \Re {\vec{\hat{q}}e^{i{\omega}\text{t}}},
    \label{eq: fext}
\end{equation}
where, $i$ is the imaginary unit, $\vec{\hat{q}}\in\mathbb{C}^{r+2c}$ is the Fourier coefficient of the harmonic function, and $\Re{\square}$ selects the real part of a complex vector. We then seek a steady-state solution to Eq. \ref{eq: EOM}. Such a solution can be derived either in the time or the frequency domain. Regarding the solution in the frequency domain, the Multi Harmonic Balance (MHB) is adopted. It is worth mentioning that special attention should be paid to the nonlinear crack face forces due to their non-smooth nature. To the best of the author's knowledge, there is no rigorous mathematical proof for the existence of a steady-state solution for a continuous (i.e. $C^0$) piece-wise linear dynamical system. However, in cases where the piece-wise linear force is small (in our case, implying a small crack), a steady-state response is observed in time-domain simulations as well as experiments. In this research, the MHB results are verified by initially performing a convergence analysis to establish the number of harmonics included in the MHB Ansatz. Following this, the system is solved in the time domain for selected frequencies via direct time integration to allow for comparison. The generalized-alpha method \cite{chung1993time} was employed to avoid so-called spurious oscillations \cite{wriggers2006computational,laursen2013computational}, which are a numerical artifact due to the non-smoothness of the nonlinear forces.

Let us assume that  $\vec{x}\text{(t)}$ is the steady-state response to \ref{eq: EOM}. The MHB Ansatz then writes

\begin{equation}
    \vec{x}(\text{t}) \approx \Re{\sum_{\text{p}=-h}^{h} \vec{\hat{x}}^{\text{(p)}}e^{i\text{p}{\omega}\text{t}}},
    \label{eq: xdecomp}
\end{equation}
where, $\vec{\hat{x}}^{\text{(p)}}(\omega)\in\mathbb{C}^{r+2c}$ is the Fourier coefficient for the $\text{p}^{th}$ harmonic, and $h$ is the truncated number of harmonics in the MHB Ansatz. The nonlinear force vector, $\vec{f}(\vec{x}(\text{t}))$ in Eq. \ref{eq: EOM}, can be treated as a general periodic function with period $\text{T} = \frac{2\pi}{\omega}$ and, therefore, can be decomposed into a Fourier series basis as $\vec{x}(\text{t})$:

\begin{equation}
\vec{f}(\text{t}) \approx \Re {\sum_{\text{p}=-h}^{h} \vec{\hat{f}}^{\text{(p)}}e^{i\text{p}{\omega}\text{t}}}.
\label{eq: fnldecomp}
\end{equation}
The $\text{p}^{th}$-order Fourier coefficients of the nonlinear force $\vec{\hat{f}}^{\text{(p)}}$ can be derived as:

\begin{equation}
\vec{\hat{f}}^{\text{(p)}} = \frac{1}{\text{T}}\int_{0}^{\text{T}}\vec{f}e^{-i\text{p}{\omega}\text{t}} dt, \hspace{1cm} p = -h, ..., h.
\label{eq: fouriercoefs}
\end{equation}
The transition of the contact state in the time domain renders the evaluation of the above integrals non-trivial. Therefore, the Fourier coefficients of the nonlinear force, ${\vec{\hat{f}}(\omega)}$, are derived either by analytical evaluation of the transition times \cite{petrov2003analytical}, or by the Alternating Frequency-Time (AFT) procedure \cite{cardona1994multiharmonic,cameron1989alternating}. The latter is adopted here for the convenience of extending the method to incorporate higher harmonics in the MHB Ansatz. It is observed from Eq. \ref{eq: fouriercoefs} that the non-zero higher-order harmonic terms stem from the piecewise linear force between the crack faces, leading to non-zero terms for $\text{p}>1$. 

After the Fourier decomposition, the set of $n$ differential equations of motion, Eq. \ref{eq: EOM}, transforms into a set of $(h+1)\times (r+2c)$ complex algebraic equations in the frequency domain. Substituting Eq. \ref{eq: fext}, Eq. \ref{eq: fnldecomp}, and Eq. \ref{eq: xdecomp} into Eq. \ref{eq: EOM} yields:

\begin{equation}
    \begin{bmatrix}
        \mat{D}^{(0)} & 0 & 0 & \dots & 0 \\
        0 & \mat{D}^{(1)} & 0 & \dots & 0 \\
        0 & 0 & \mat{D}^{(2)} & \dots & 0 \\
        \vdots& \vdots & \vdots & \ddots & \vdots \\
        0 & 0 & 0 & \dots & \mat{D}^{\text{(h)}} \\
    \end{bmatrix}
    \begin{bmatrix}
        \vec{\hat{x}}^{(0)}\\
        \vec{\hat{x}}^{(1)}\\
        \vec{\hat{x}}^{(2)}\\
        \vdots\\
        \vec{\hat{x}}^{\text{(h)}}\\
    \end{bmatrix}
    +\begin{bmatrix}
        \vec{\hat{f}}^{(0)}\\
        \vec{\hat{f}}^{(1)}\\
        \vec{\hat{f}}^{(2)}\\
        \vdots\\
        \vec{\hat{f}}^{\text{(h)}}\\
    \end{bmatrix}
    -\begin{bmatrix}
        \vec{0}\\
        \vec{\hat{q}}^{(1)}\\
        \vec{0}\\
        \vdots\\
        \vec{0}\\
    \end{bmatrix} = \Vec{0},
    \label{eq: dynstiffmat}
\end{equation}
where $\mat{D}_\text{p} \in \mathbb{C}^{(r+2c)\times(r+2c)} $ is the $\text{p}^{th}$-order complex dynamic stiffness matrix defined as:

\begin{equation}
    \mat{D}^{(\text{p})} := -(\text{p}\omega)^2\mat{M} + i\text{p}\omega\mat{C} + \mat{K},  \quad \text{p} = 0,1,2,...,\text{h}.
\end{equation}
The set of nonlinear algebraic equations Eq. \ref{eq: dynstiffmat}, is then solved by an iterative Newton-Raphson technique as

\begin{equation}
    \vec{R}(\hat{\vec{x}}_{\text{s}}) := \mat{D}\hat{\vec{x}}_{\text{s}} + \vec{\hat{f}}(\hat{\vec{x}}_{\text{s}}) - \vec{\hat{q}},
\end{equation}
where the residual, $\vec{R}$ is expanded in a Taylor series around a reference point as:

\begin{equation}
    \vec{R}(\hat{\vec{x}}_{\text{s+1}}) \approx \vec{R}(\hat{\vec{x}}_{\text{s}}) + \frac{\partial\vec{R}}{\partial\vec{\hat{x}}}\bigg|_{\vec{\hat{x}}_{\text{s}}}(\hat{\vec{x}}_{\text{s+1}} - \hat{\vec{x}}_{\text{s}}) = \vec{0}.
    \label{Eq: jacobian}
\end{equation}
In Eq. \ref{Eq: jacobian}, the subscript $\text{s}$ denotes the iteration number. The partial derivative of the residual with respect to $\vec{\hat{x}}$, indicated as $\frac{\partial\vec{R}}{\partial\vec{\hat{x}}}$, is referred to as the Jacobian matrix. The initial guess for solving Eq. \ref{Eq: jacobian} is obtained from the linear solution, i.e. $\vec{\hat{f}} = \vec{0}$.

\subsection{The proposed approximated solution}

\ 

This study is primarily concerned with detecting small-sized cracks, particularly those in the early stages. Thus, it is assumed that the magnitude of the crack face force is minor when compared to the magnitude of the internal forces, specifically $\norm{\vec{f}} << \norm{\mat{M}\ddot{\vec{x}} + \mat{C}\dot{\vec{x}} + \mat{K}\vec{x}}$. Consequently, a first-order approximation to Eq. \ref{eq: EOM} can be obtained by representing the \emph{breathing} crack model as a small perturbation of the \emph{pristine} model. To implement the former within a finite element setting, we begin by utilizing the Lagrange Multipliers approach to model the \emph{pristine} scenario. This involves imposing the compatibility constraint, specifically $\Delta\vec{x}_\text{n} = 0$, on the surfaces of the crack in a dual manner (refer to \cite{geradin2015mechanical}).

\begin{equation}
\begin{split}
     &\mat{M}\ddot{\vec{x}}\text{(t)} + \mat{C}\dot{\vec{x}}\text{(t)} + \mat{K}\vec{x}\text{(t)} + \mat{B}^{T}\vec{\lambda} = \vec{q}\text{(t)}, \\
     &\mat{B}\vec{x}\text{(t)} = \vec{0}.
     \label{eq:EOM closed}
     \end{split}
\end{equation}

where $\mat{B} \in\mathbb{R}^{c\times(2c+r)}$ is the signed Boolean matrix as defined before, $\lambda \in\mathbb{R}^{c}$ is the force required to ensure compatibility at the crack faces (i.e. $\Delta \vec{x} = 0$).

Adding $\vec{\lambda}$ to the set of unknowns yields:

\begin{equation}
    \mat{\overline{M}}\ddot{\vec{\overline{x}}}+ \mat{\overline{C}}\dot{\vec{\overline{x}}} + \mat{\overline{K}}\vec{\overline{x}} = \vec{\overline{q}},
\end{equation}
where,
\begin{equation}
    \mat{\overline{M}} = \left[
    \begin{array}{cc}
        \mat{M} & \mat{0} \\
        \mat{0} & \mat{0} \\
    \end{array}
    \right],\hspace{0.5cm} \mat{\overline{C}} = \left[
    \begin{array}{cc}
        \mat{C} & \mat{0} \\
        \mat{0} & \mat{0} \\
    \end{array}
    \right],\hspace{0.5cm} \mat{\overline{K}} = \left[
    \begin{array}{cc}
        \mat{K} & \mat{B}^{T} \\
        \mat{B} & \mat{0} \\
    \end{array}
    \right],\hspace{0.5cm} \vec{\overline{x}} = \left[
    \begin{array}{c}
         \vec{x} \\
         \vec{\lambda} \\
    \end{array}
    \right],\hspace{0.5cm} \vec{\overline{q}} = \left[
    \begin{array}{c}
         \vec{q} \\
         \vec{0} \\
    \end{array}
    \right].       
     \label{eq:EOM_reathing}
\end{equation}

The above equation of motion pertains to the \emph{pristine} case (i.e. the crack remains closed at all times). Then, the EoM for the case of a \emph{Breathing} crack can be formulated as:

\begin{equation}
    \mat{\overline{M}}\ddot{\vec{\overline{x}}}+ \mat{\overline{C}}\dot{\vec{\overline{x}}} + (\mat{\overline{K}} - \epsilon\Delta\mat{\overline{K}}(\vec{\lambda}))\vec{\overline{x}} = \vec{\overline{q}}.
    \label{eq: EOM_approx}
\end{equation}
In Equation \ref{eq: EOM_approx}, $\epsilon$ represents a small positive scalar ($0<\epsilon<<1$) that implies the crack size is minimal, resulting in only a minor reduction in the overall stiffness due to the crack. Then $\Delta\mat{\overline{K}}$ can be written as:
\begin{equation}
\Delta\mat{\overline{K}} = 
\left[
    \begin{array}{cc}
        \mat{0} & \mat{B}^T \text{diag}(H(\vec{\lambda})) \\
        \text{diag}(H(\vec{\lambda}))\mat{B} & \mat{0} \\
    \end{array}
    \right],
\end{equation}
where, $H(\square)$, as defined previously, represents the step function that regulates the breathing mechanism of the crack. Essentially, crack face forces are removed when in tension to allow the separation of crack faces, whereas they are maintained when in compression to prevent the surfaces from penetrating into each other.

We then expand $\vec{\overline{x}}$ following the perturbation method in the form:

\begin{equation}
     \vec{\overline{x}}(\text{t};\epsilon) = \vec{\overline{x}}_{0} (\text{t}) + \epsilon \vec{\overline{x}}_{1}(\text{t}) + O(\epsilon^2) = \left[
    \begin{array}{c}
         \vec{x}_0 + \epsilon \vec{x}_1 + O(\epsilon^2)\\
         \vec{\lambda}_0 + \epsilon\vec{\lambda}_1 + O(\epsilon^2)\\
    \end{array}
    \right],
     \label{eq: expansion}
\end{equation}
Substituting Eq. \ref{eq: expansion} into Eq. \ref{eq: EOM_approx}, and setting the coefficients of each power of $\epsilon$ independently to zero, results in the following:

\begin{equation}
    \epsilon^0 \rightarrow
    \mat{\overline{M}}\ddot{\vec{\overline{x}}}_0+ \mat{\overline{C}}\dot{\vec{\overline{x}}}_0 + \mat{\overline{K}}\vec{\overline{x}}_0 = \vec{\overline{q}},
     \label{eq: order 0}
\end{equation}

\begin{equation}
    \epsilon^1 \rightarrow
    \mat{\overline{M}}\ddot{\vec{\overline{x}}}_1+ \mat{\overline{C}}\dot{\vec{\overline{x}}}_1 + \mat{\overline{K}}\vec{\overline{x}}_1 = \Delta\mat{\overline{K}}(\vec{\lambda}_0)\vec{\overline{x}}_0.
     \label{eq: order 1}
\end{equation}
Where Eq. \ref{eq: order 0} ($\epsilon^0$) corresponds to the \emph{pristine} case. In Eq. \ref{eq: order 1}, $\vec{\lambda}_0$ are the Lagrange multipliers calculated for the \emph{pristine} case, and $\vec{\lambda}_1$ denotes the deviation in the internal forces due to the crack. One can show that the effect of $\vec{\lambda}_1$ in Eq. \ref{eq: order 1} is of the order $\epsilon$ and can be neglected.
\begin{equation}
    H(\vec{\lambda}_0+\epsilon \vec{\lambda}_1) = H(\vec{\lambda}_0) + \epsilon\delta(\vec{\lambda}_0)\vec{\lambda}_1 = H(\vec{\lambda}_0) + O(\epsilon),
     \label{eq: heaviside approx}
\end{equation}
Thus, this results in Eq. \ref{eq: order 1}.

Finally, since the RHS of \ref{eq: order 1} is a periodic function with frequency $\omega$, its solution in the frequency domain is obtained by approximating the forcing vector using a sum of harmonic functions via the Fourier transform, as done in \ref{eq: fouriercoefs} as:

\begin{equation}
\Delta\mat{\overline{K}}\vec{\overline{x}}_0 \approx \Re{\sum_{\text{p}=-h}^{h} \vec{\hat{F}}^{\text{(p)}}e^{i\text{p}{\omega}\text{t}}}.
\label{eq: fnldecomp approx}
\end{equation}
where the Fourier coefficients of the nonlinear force vector $\vec{\hat{F}}^{\text{(p)}}$, in Eq. \ref{eq: fnldecomp approx}, can be obtained as:

\begin{equation}
     \vec{\hat{F}}^{\text{(p)}} \approx \frac{\omega}{2\pi}\int_{0}^{\frac{2\pi}{\omega}}\Delta\mat{\overline{K}}(\vec{\lambda}_0)\vec{\overline{x}}_0 e^{-i\text{p}{\omega}\text{t}} dt.
     \label{Eq: fouriercoefs_approx}
\end{equation}
Finally, the overall solution becomes as follows:

\begin{equation}
    \vec{\overline{x}}(\text{t};\epsilon) = \Re{\hat{\vec{\overline{x}}}_0^{(1)}e^{i\omega \text{t}} + \hat{\vec{\overline{x}}}_0^{(-1)}e^{-i\omega \text{t}} + \epsilon\sum_{\text{p}=-h}^{h}\hat{\vec{\overline{x}}}^{(\text{p})}_1e^{i\text{p}{\omega}\text{t}}},
 \label{Eq: 1 order solution}
\end{equation}
where,
\begin{equation}
\hat{\vec{\overline{x}}}_0^{(1)} = [\mat{\overline{D}}]^{-1}\vec{\overline{q}}, \hspace{1cm} \hat{\vec{\overline{x}}}^{(\text{p})}_1 = [\mat{\overline{D}}^{(\text{p})}]^{-1}\vec{\hat{F}}^{\text{(p)}}.
 \label{Eq: fourier coe approx}
\end{equation}

Using the proposed method, the iterative solution is eliminated for evaluating the nonlinear response. Calculating $\vec{\lambda}$ only requires the solution for the pristine axle (through Eq. \ref{eq: order 0}), which can be calculated off-line and only once for all crack scenarios. This will bring a significant speed gain in solving the inverse problem. The proposed method can substantially accelerate the simulated response process and ultimately facilitate in-service crack identification, as discussed in the following sections.
\subsection{Crack identification setup using second-order transmissibility}
\label{section: 2.5}
In its more general form, transmissibility is defined as a frequency-dependent ratio of responses (e.g., displacement, velocity, or acceleration) between two points in a structure or system \cite{ribeiro2000generalisation}. Transmissibility functions were first devised for the periodically forced vibration of linear stable dynamical systems and have been successfully implemented in SHM \cite{Agathos2021Parametric, MAIA20112475}. The main advantage of using transmissibility functions for SHM lies in the fact that these functions do not depend on the magnitude of the external force. This is of great relevance for in-service monitoring, as the external loading changes continuously during operation and is therefore not easily monitored. However, the insensitivity of the linear features (including transmissibility functions) for the identification of the incipient cracks renders the methods based on conventional transmissibility ineffective \cite{sinou2009review,Hiwarkar2012,lin2018higher}. As a remedy for the aforementioned challenge, we define the transmissibility of higher-order harmonics, referred to here as HOTr, as a DSF that is tailored for crack detection.

We define $\text{p}^{th}$-order transmissibility $\hat{\text{Tr}}^\text{(p)}_\text{m,n} \in\mathbb{C}^{1}$, between the $\text{p}^{th}$-order frequency domain response at $\text{m}^{th}$ DoF $\hat{\text{x}}^\text{(p)}_\text{m} \in\mathbb{C}^{1}$, and $\text{n}^{th}$ DoF $\hat{\text{x}}^\text{(p)}_\text{n} \in\mathbb{C}^{1}$, as: 

\begin{equation}   \hat{\text{Tr}}^\text{(p)}_\text{m,n} := \frac{\hat{\text{x}}^\text{(p)}_\text{m}}{\hat{\text{x}}^\text{(p)}_\text{n}}.
     \label{eq:FEDynEquil}
\end{equation}
As shown in the previous section, the set of nonlinear algebraic equations, reflected in Eq. \ref{eq: dynstiffmat}, must be solved to calculate $\vec{\hat{x}}^{\text{(p)}}$. Since this is computationally expensive, the solution to the inverse problem associated with crack parameter identification can be severely hampered. Therefore, here we use the first-order approximated solution provided in the previous section to calculate the HOTr.

Let us now assume that a crack can be defined through a set of parameters contained in the entries of the vector $\vec{\theta} \in \mathbb{R}^{v}$, where $v$ represents the dimension of the parameter space. Then, the crack identification procedure can be set through an inverse problem formulated as an optimization problem, seeking to find the optimal variable of the parameter vector $\vec{\theta}$, as follows:

\begin{equation}
    \vec{\tilde{\theta}} \simeq \text{argmin}_{\vec{\theta} \in \vec{\Theta}} J_{\vec{\theta}}.
    \label{Eq. argmin}
\end{equation}
where, $\vec{\Theta}$ is the set of all possible values of $\vec{\theta}$, and $J_{\vec{\theta}}$ is a regularized metric for the difference between the measured and simulated DSF. As mentioned above, in this paper, we suggest employing \emph{second-order} transmissibility as the DSF, thereby defining $\vec{\eta}$ as the crack identification feature, which is a vector of all possible second-order transmissibilities:
\begin{equation}
     \vec{\eta} := \text{Tr}_{m,n}^{(2)}, \hspace{0.5cm} \{m,n\}\in \binom{Q}{2},
     \label{Eq. damage sensitive feature}
\end{equation}
where $Q$ is the set of all sensors, and indices $m$ and $n$ are chosen as an unordered pair from the set $Q$. The objective function for crack identification is then defined as:
\begin{equation}
     J_{\vec{\theta}} := \frac{\lVert\vec{\eta}_{s}(\vec{\theta}) - \vec{\eta}_{m}\lVert} {\lVert\vec{\eta}_{m}\lVert}\times 100.
     \label{Eq. objective function}
\end{equation}
In Eq. \ref{Eq. objective function}, $\vec{\eta}_{s}(\vec{\theta})$ and $\vec{\eta}_{m}$ denote the simulated and measured crack identification features.

The extraction of higher-order harmonics from the time-domain measurements is achieved by a recursive correlation method proposed in \cite{Lin2018} that follows:

\begin{equation}
\vec{\hat{x}}^{\text{(p)}} = \frac{1}{\text{T}_s}\int_{\text{0}}^{\text{T}_s} (\vec{x} - \sum_{j = 1}^{\text{p}-1} \vec{x}^{(j)}) e^{-ip{\omega}\text{t}} d\text{t},
\label{Eq. jthHRMinfreqcoef}
\end{equation}
where $\vec{x}^{(j)}(\text{t}) \in\mathbb{R}^{n} $ is the $j^{th}$ harmonic in the time domain, given by

\begin{equation}
\vec{x}^{(j)}(\text{t}) = \Re{\vec{\hat{x}}^{(j)}e^{i\text{j}{\omega}\text{t}}}, \quad j = 1,2,...,\text{h}.
\label{Eq. jthHRMintime}
\end{equation}
Note that the dependency of the $\vec{x}$ to 
$\text{t}$ is dropped in the right-hand side of Eq. \ref{Eq. jthHRMinfreqcoef} for the sake of convenience. The method effectively extracts the higher-order harmonics for a specific $\omega$ by minimizing the leakage and aliasing effects \cite{Lin2018}. In practice, the same data acquisition system used for conventional vibration monitoring is sufficient to capture higher-order harmonics; only an appropriately increased sampling rate is required to ensure that the relevant frequency content is resolved.

The solution to the optimization problem can be realized through diverse schemes. In our case, a closed-form dependency relationship between the objective function and the parameters is not available. In addition, the objective function is highly non-convex and non-smooth. Therefore, we opt to apply gradient-free optimization methods here. Examples of methods that are often adopted within inverse problem settings in an SHM context include Genetic Algorithms (GA) \cite{rabinovich2007xfem,waisman2010detection}, Particle Swarm Optimization (PSO) \cite{kennedy1995particle,Agathos2021Parametric}, Artificial Bee Colony (ABC) \cite{YANG2015219,sun2013nondestructive}, as well as the Covariance Matrix Adaptation Evolution Strategy (CMAES) \cite{hansen2006cma,agathos2018multiple,agathos2021crack}. In this work, we adopt the Genetic Algorithm (GA) as one potential approach, acknowledging the implications of the No Free Lunch Theorem. In terms of implementation, MATLAB's built-in GA function is used. The hyper-parameters of the method, such as the population size and maximum number of generations, are chosen according to the specific parameter space for the numerical examples at hand; see Section \ref{section: numerical examples}. In this setup, the crossover coefficient, representing the proportion of the next generation (excluding elite children) generated by crossover, is fixed at 0.8. Additionally, a Gaussian distribution for mutation is employed with both scale and shrink parameters equal to 1. Default values are used for the rest of the parameters.

The solution to the optimization problem using the GA algorithm is achieved by repeatedly solving the forward problem (i.e., simulating the response of the structure with a given $\vec{\theta}$). The procedure followed for solving the inverse problem is given in Algorithm \ref{alg: GA}.

It should be noted that for each parameter set $\vec{\theta}$, the corresponding FE model needs to be constructed and then solved (steps 4 and 5 in the algorithm). The challenge of reducing the computational cost associated with the solution step (step 5) is addressed by the proposed method of approximating HOTr. For the construction of the FE model (step 4), the substructuring technique is used, as discussed in the next section.

The measured HOTr is synthetically simulated using a time integration procedure. The generalized alpha method \cite{arnold2007convergence} with a fixed time step is implemented for this purpose. The high-frequency numerical damping coefficient, $\rho_{\infty}$, is set to 0.7 after conducting a relevant convergence analysis. In mimicking realistic measurements, the simulated measured signals are corrupted with additive white noise, which is generated independently for each sensor reading.

\begin{equation}
    \text{S}_{m}(\text{t}) = \text{S}(\text{t}) + e(\text{t}),
\end{equation}
where $\text{S}(\text{t})$ is a quantity of interest and $\text{S}_{m}(\text{t})$ is the corresponding noisy measurement, with $e(\text{t})$ being the white noise added to the signal. Three different noise levels are considered: 1\%, 5\%, and 10\% Root Mean Square (RMS) noise to signal ratio, defined as

\begin{equation}
\text{noise level} = \sqrt{\frac{\int_{\text{0}}^{\text{T}_s}e^{2}(\text{t})}{\int_{\text{0}}^{\text{T}_s}\text{S}^{2}(\text{t})}d\text{t}}\times100
\end{equation}
\begin{algorithm}[h]
\caption{Crack identification using GA}
\label{alg: GA}
  \begin{algorithmic}[1]
  \REQUIRE  Finite Element mesh of the pristine model (i.e. without crack), $\vec{\Theta}, {_{m}}\text{Tr}_{m,n}^{(2)}$ 
  \ENSURE  $\vec{\tilde{\theta}}$
  \WHILE {min($J_{\vec{\Theta_{b}}}$)$>$ threshold}
  \STATE determine $\Theta_{b} \subset \Theta^{\textcolor{red}{a}}$ 
  \FOR{$\theta \in \Theta_{b}$}
  \STATE construct the FE model with crack defined by $\vec{\theta}$ (Eq. \ref{eq: EOM})
  \STATE calculate the proposed approximation to $_s\hat{\text{Tr}}_{m,n}$
  \ENDFOR
  
  \STATE evaluate $J_{\vec{\theta}}$ for all $\vec{\theta} \in \vec{\Theta_{b}}$ (Eq. \ref{Eq. objective function})
  \ENDWHILE
\end{algorithmic}
\hrulefill \\
$^{\textcolor{red}{a}}$ selection of the $\Theta_{b}$ out of $\Theta$ is carried out by GA algorithm see \cite{goldberg1989genetic} for details.
\end{algorithm}

\section{Numerical examples}
\label{section: numerical examples}

In this section, we demonstrate the concepts previously introduced using numerical examples. Specifically, we consider two-dimensional and three-dimensional finite element models of a train wheelset axle. Initially, we examine the sensitivity of the second-order harmonic response to crack parameters, illustrating the impact of varying crack positions on this inferable quantity. Next, we evaluate the proposed method for approximated HOTrs, focusing on accuracy and computational speedup. Finally, we implement the second-order transmissibility of strain readings for crack identification and assess the capability of the proposed approach via Monte Carlo analysis.

To facilitate the solution of the derived equations, two different reduced-order models are considered here. The first model, referred to as "RB," is derived by retaining crack interface DoFs and external forcing DoFs while reducing all other DoFs via the Rubin method \cite{rubin1975improved}. The other reduced-order model, denoted here as "SUB," is obtained by substructuring. The idea is to divide the structure into two possible substructures: substructure A, which is the area of the structure where possible cracks are expected to form, and substructure B, which contains the remaining domain (see Figure \ref{fig: 3models - 2D} and Figure \ref{fig:3model - 3D} for a graphical rendition of the 2D and 3D models, respectively). It should be noted that no specific assumption, other than the fact that cracks would form on the surface of the wheelset axle, is made regarding the selection of substructure A. Then, substructure A is kept as a full-order model, while substructure B is reduced via the Rubin method, keeping only the DoFs at the interface with substructure A and the external forcing DoFs. The two substructures are then assembled by imposing compatibility of the displacements at their interface. The resulting RB model is smaller in terms of DoFs and is thus faster for a single forward evaluation. The main advantage of using the SUB model over the RB model is that the construction of substructure B can be done offline (once for all crack cases). This drastically reduces the overall online cost, as shown in the following sections. The FE implementation is performed on the MATLAB-based YAFEC code \cite{shobhitjainyafec}.

\subsection{2D example}

\begin{figure}[h]
    \centering
    \adjincludegraphics[width=6in,trim={0 {0.26\height} 0 0 },clip]{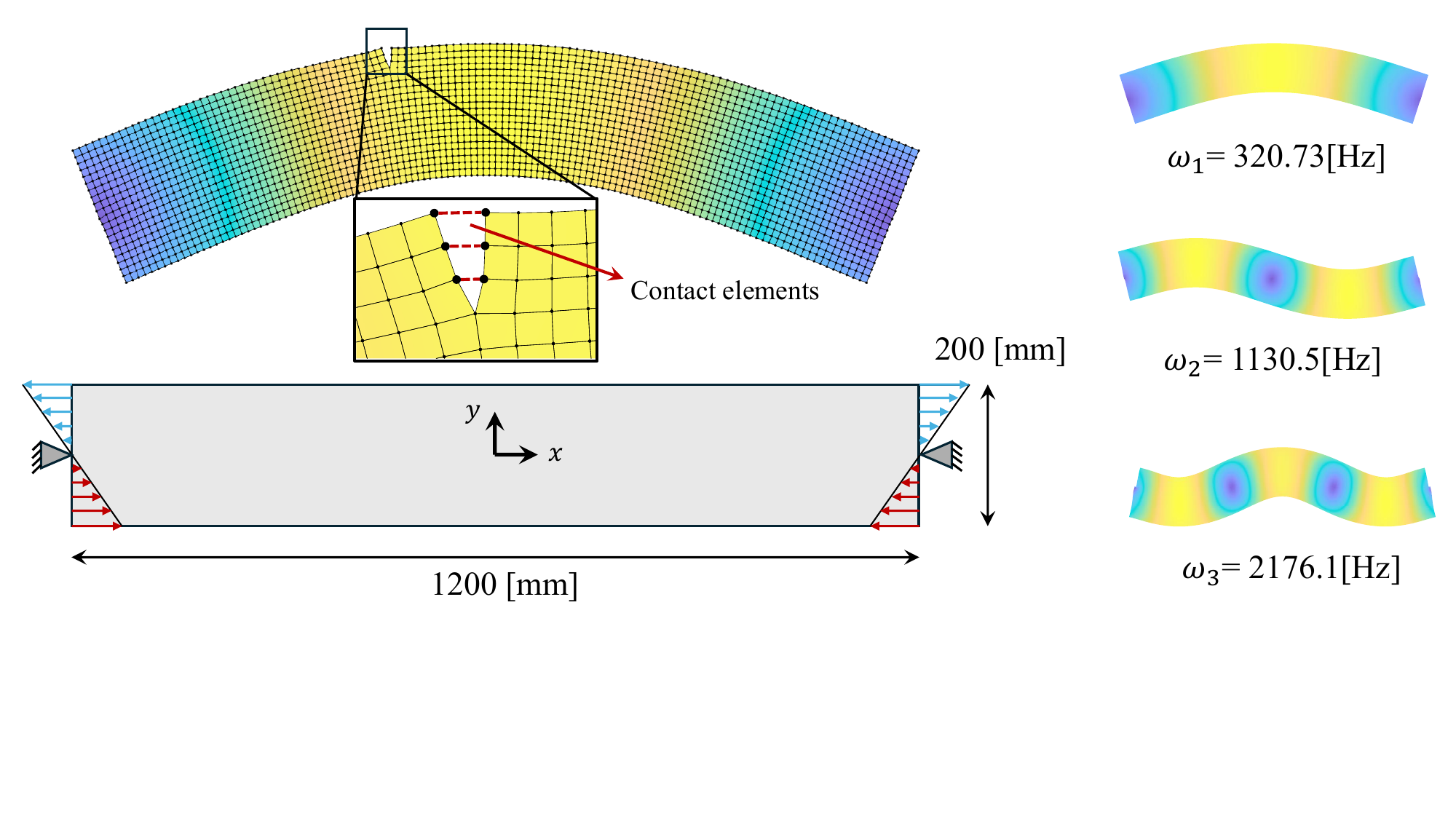}
    \caption{Example of the 2D beam: External forces, boundary conditions, and dimensions. First three vibration modes with their natural frequencies. The out pf plane thickness of the beam is 1 mm.}
    \label{fig: 2D_model}
\end{figure}

For the 2D case, we consider a rectangular beam, as shown in Figure \ref{fig: 2D_model}. Steel material characteristics are adopted, assuming a Young's modulus of $E = 2.1\times10^{5}$ [Mpa], a density of $\rho = 7.3\times10^{3}$ [Kg/$\text{m}^{3}$], and a Poisson ratio of $\nu = 0.26$. The beam is discretized using linear solid continuum elements, resulting in a total of 2400 elements and 2541 nodes. The beam is simply supported at both ends by fixing one node on each side, and bending moments of equal magnitude are realized with a linear distribution of horizontal forces acting at the ends, as shown in Figure \ref{fig: 2D_model}. The boundary conditions and the forcing are designed to replicate the inner section of the train axle undergoing rotational four-point bending. Rayleigh damping, corresponding to the damping ratio $\zeta = 0.002$, calculated for the first two bending modes, was considered. The first three vibration modes of the model are depicted in Figure \ref{fig: 2D_model}. 

The crack is characterized by two parameters: $L_c \in (-1,1)$, representing the normalized position of the crack along the beam's length, and $D_c \in\{5,10,15\}$, which indicates the percentage ratio of the crack depth to the beam width. For instance, $D_c = 5$ corresponds to a crack depth of 5\% of the thickness of the beam. The crack's geometry must align with the boundaries of the elements within the model, resulting in a discrete parameter space for the crack described by $\vec{\Theta}\in \mathbb{R}^{120\times3}$. The breathing crack is modeled as a frictionless contact between each pair of contact nodes, as outlined in section \ref{section: Methods}(see Fig. \ref{fig: 2D_model}). 

\begin{figure}[h]
    \centering
    \adjincludegraphics[width=6in,trim={0 {0.72\height} {0.13\width} 0 },clip]{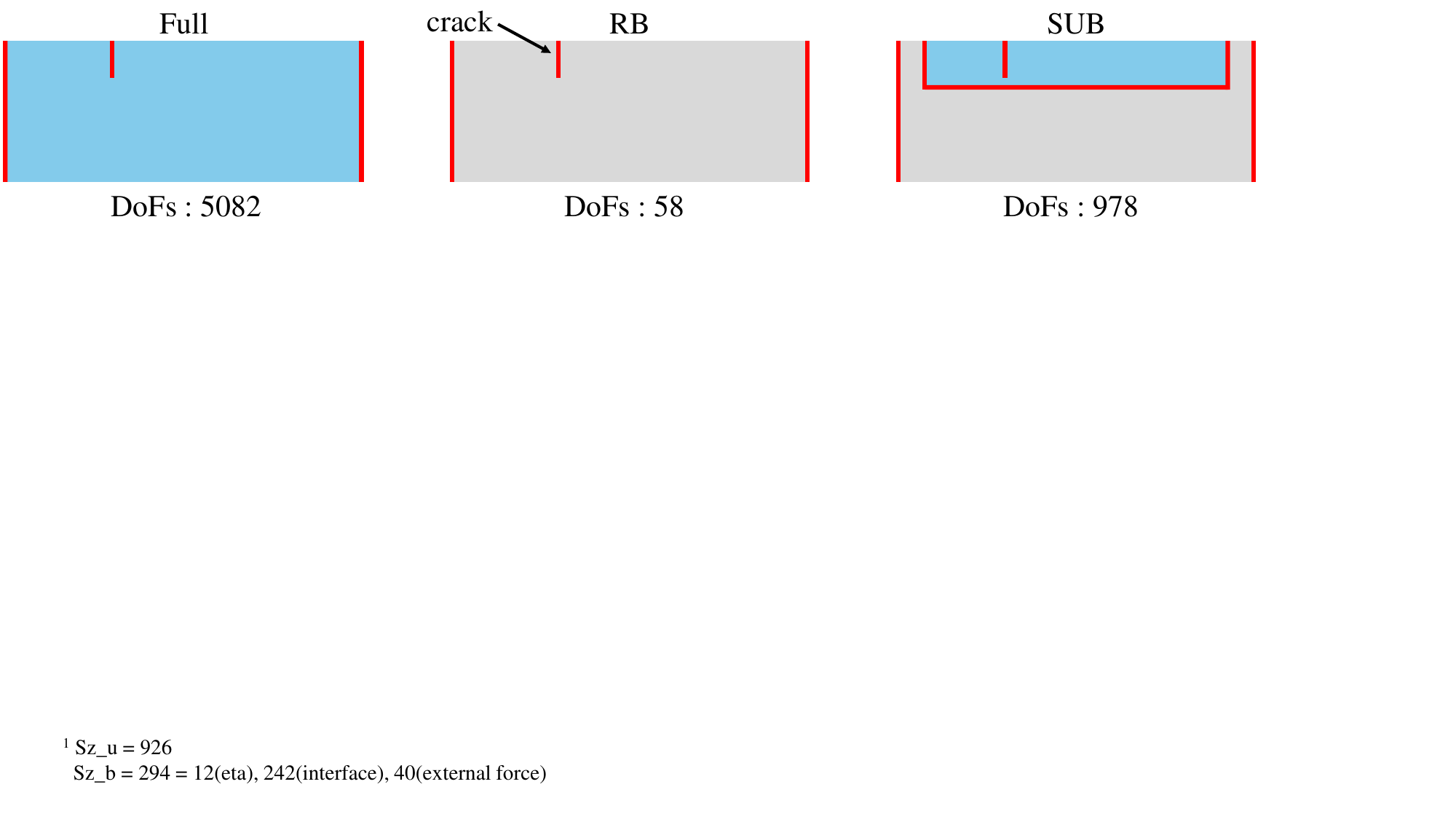}
    \caption{Schematics of full-order model of the 2D beam and two reduced-order models considered, RB and SUB.}
    \label{fig: 3models - 2D}
\end{figure}

Next, two reduced-order models, namely RB and SUB, were derived. For the RB model, 6 vibration modes are used for the reduction basis of the Rubin method. For the SUB model, the area near the top edge of the beam was chosen to be substructure A, which encompasses all the possible crack scenarios ($\Theta$). Substructure B, corresponding to the remaining portion of the beam, is then reduced using the Rubin method by considering 6 vibration modes in the reduction basis. The two models and resulting DoFs are shown in Figure \ref{fig: 3models - 2D}.

\subsubsection{Sensitivity of HOTr to the crack parameters}

In this section, we examine the sensitivity of the second-order harmonic displacement ($\hat{\vec{x}}^{\text{(2)}}$) to the crack position. For this analysis, the results were obtained from nonlinear simulations using the MHB method for a specified frequency. A convergence analysis is conducted to determine the number of harmonics to be considered, the step size, and the number of discretization points per cycle for the AFT algorithm due to the non-smooth behavior of the breathing crack. The frequency of the employed excitation was set to 128 Hz to resemble a real-world scenario in which the excitation frequency is below the axle's first natural frequency. The amplitude of the force was chosen to ensure that, under static conditions, the deformation reached a maximum of 2 mm at the midpoint of the beam (equivalent to 1\% of its width). The crack depth was set to 10\% of the beam's thickness ($D_c=10$).

\begin{figure}[h]
    \centering
    \adjincludegraphics[width=7in,trim={0 0 0 0 },clip]{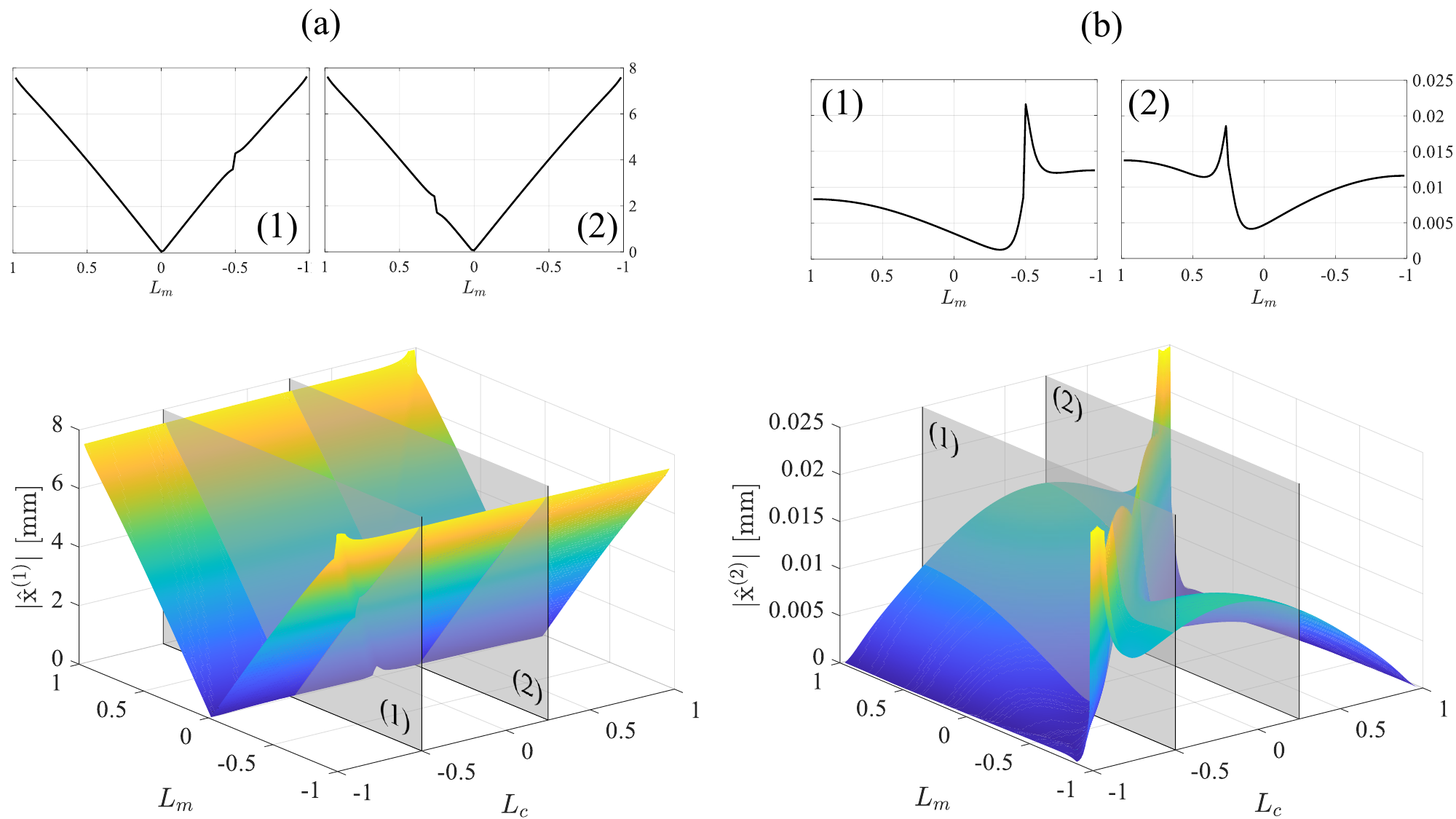}
    \caption{(a) First-order displacement response $\hat{\vec{x}}^{\text{(1)}}$, and (b) second-order displacement response $\hat{\vec{x}}^{\text{(2)}}$ along the surface of the beam against varying location of the crack ($L_c$). The measurement point is denoted as $L_m$.}
    \label{fig: FRF_2D}
\end{figure}

To observe how the second-order response along the axle varies with different crack locations, Figure \ref{fig: FRF_2D}(b) illustrates the \emph{second-order} displacement in $x$ direction ($\hat{\vec{x}}^{\text{(2)}}$) at the upper structure boundary, plotted against the crack location. The second-order harmonic displacement is observed to be an attribute that is sensitive to local defects, as its pattern across the surface significantly changes in response to shifts in crack location. Such sensitivity is crucial for the successful implementation of damage localization schemes. At the same time, $\hat{\vec{x}}^{\text{(2)}}$ does not vanish from the crack location, implying that a potential crack can be detected on the basis of the second-order harmonics from a sparse measurement grid, which again is an essential trait for cost-efficient SHM schemes. For comparison, first-order harmonic displacement in the x direction is plotted in Figure \ref{fig: FRF_2D}(a). Aligned with previous reports in the literature (e.g. \cite{sinou2009review,lin2018higher}), it is observed that the second-order harmonic response shows substantially higher sensitivity to variation in crack location with respect to its first-order counterpart.

\subsubsection{Approximation of the HOTrs}

\begin{figure}[h]
    \centering
    \adjincludegraphics[width=7in,trim={0 {0.245\height} {0.18\width} 0 },clip]{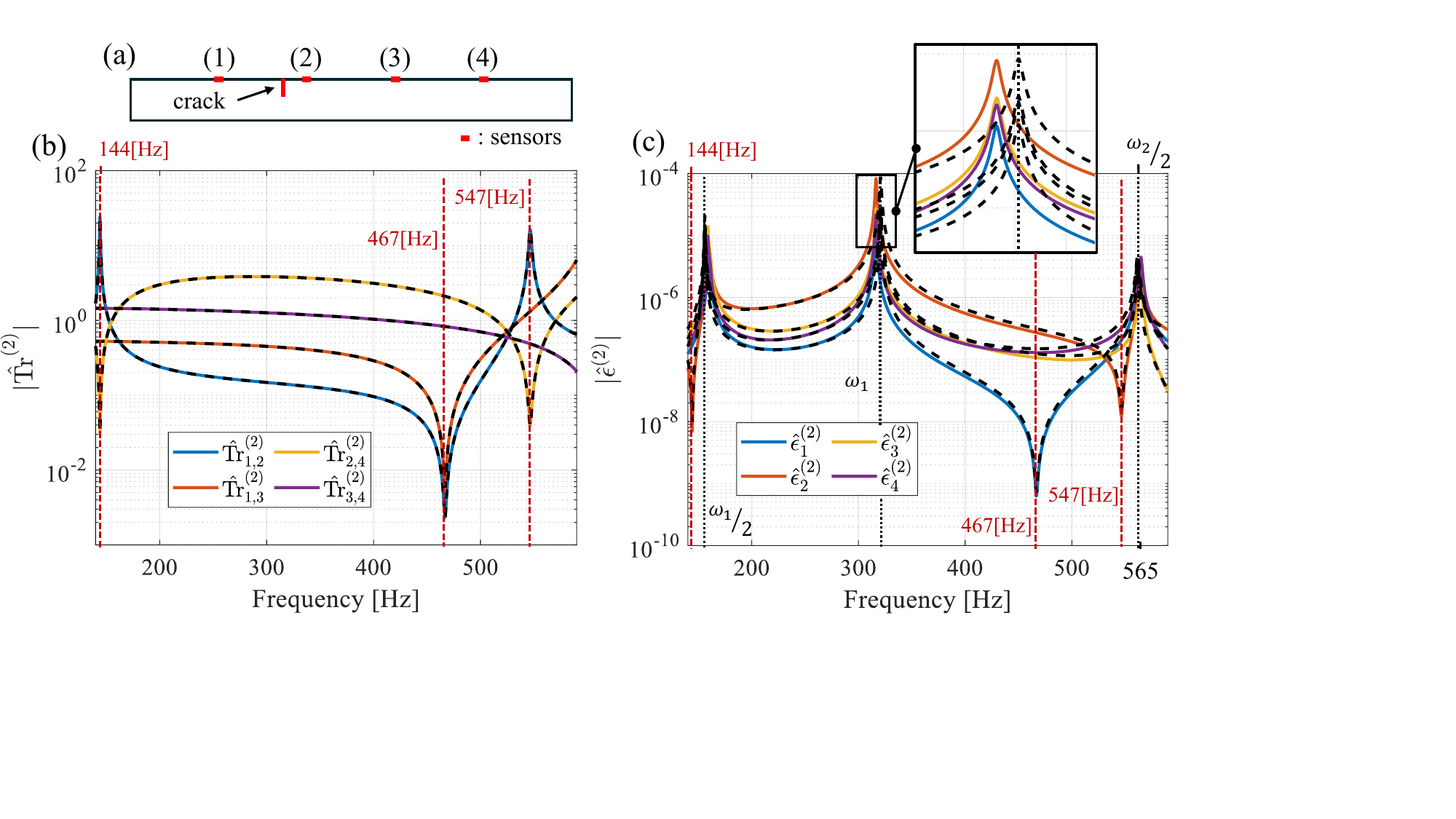}
    \caption{(a) Schematics of the beam, crack, and sensor locations. (b) Second-order transmissibility of strain readings; solid lines represent nonlinear solution obtained with MHB, dashed lines the approximation using the proposed method. Maximum and average Root Mean Squared Error (RMSE) for all four curves are: $3.804\times10^{-4}$, and $1.356\times10^{-5}$ respectively. (b) Second-order strains at four sensor locations; solid lines represent nonlinear solution obtained with MHB, dashed lines the approximation using the proposed method.}
    \label{fig:TR_2D}
\end{figure}

After proving HOTr to be an effective DSF, the performance of the proposed method for approximating the second-order transmissibility is tested here. To this end, a crack with $D_c = 15$, and $L_c = -0.1667$ is considered, and $\hat{\text{Tr}}^{(2)}$ is calculated using the second-order \emph{strain} reading in the $x$ direction according to Figure \ref{fig:TR_2D}, which comes from four different virtual strain gauges that are placed equidistantly on the beam (see Figure \ref{fig: FRF_2D}(a)). The second-order strains are denoted as $\hat{\epsilon}_{j}^{(2)}$, where the superscript $j\in \{1,2,3,4\}$ corresponds to the location of the virtual strain gauge according to Figure \ref{fig:TR_2D}(a). Both the nonlinear solution and the approximation were carried out on the RB reduced-order model.

The results illustrated in Figure \ref{fig:TR_2D}(b) show an excellent match between the nonlinear solution of $\hat{\text{Tr}}^{(2)}$ and its approximated counterpart, which is computed using the proposed method. It should be mentioned that the peaks and anti-peaks of $\hat{\text{Tr}}^{(2)}$ in Figure \ref{fig:TR_2D}(b) are due to the \emph{anti-resonances} of $\hat{\epsilon}^{(2)}$. To delve more into the details of these peaks, $\hat{\epsilon}^{(2)}$ at four sensors are plotted in Figure \ref{fig:TR_2D}(c). As observed from the figure, $\hat{\epsilon}^{(2)}_1$ comprises an anti-resonance at approximately 467 Hz. This gives rise to the anti-peaks in the curves $\hat{\text{Tr}}^{(2)}_{1,2}$ and $\hat{\text{Tr}}^{(2)}_{1,3}$ since $\hat{\epsilon}^{(2)}_1$ remains in the numerator of the transmissibility function. Similarly, $\hat{\epsilon}^{(2)}_2$ has two anti-resonances, one at 144 Hz and the other at 547 Hz. This is evident by observing the peaks of the $\hat{\text{Tr}}^{(2)}_{1,2}$ curve in Figure \ref{fig:TR_2D}(b), where $\hat{\epsilon}^{(2)}_2$ is in the denominator, and the anti-peaks in $\hat{\text{Tr}}^{(2)}_{2,4}$, where $\hat{\epsilon}^{(2)}_2$ is in the numerator. Moreover, three resonances are observed for all the points in Figure \ref{fig:TR_2D}(c), marked with black dotted lines. The one at $\omega_1 = 316$ Hz corresponds to the first natural frequency of the beam, and the other two at $\omega_2/2 = 158$ Hz and $\omega_3/2 = 565$ Hz correspond to the sub-harmonics of the first and second natural frequencies, respectively. The decrease in the resonance frequencies compared to those from the pristine beam, shown in Figure \ref{fig: 2D_model} denotes the stiffness reduction due to the crack. As these peaks are uniform across all four points, they do not induce peaks or anti-peaks in $\hat{\text{Tr}}^{(2)}$. As illustrated in Figure \ref{fig:TR_2D}(c), there is a subtle decrease in the nonlinear resonance frequency compared to its estimated solution. However, the second-order harmonic derived through the proposed method exhibits remarkable consistency with the solutions obtained from nonlinear analysis.

The time associated with model construction and the solution for a single cracked model simulation is given in Table \ref{Tabel 1}. For this specific test, the excitation frequency was 128 Hz. The simulations are carried out on a local computer with 128 GB of RAM and an Intel(R) Xeon(R) W-2245 CPU @ 3.90 GHz processor.  The speedup gained by using the proposed approximation method as opposed to the calculation of the nonlinear solution is depicted in Table \ref{Tabel 1}.

\begin{table}[h]
\caption{Computational time comparison between the nonlinear solution (NL), and proposed approximation (APX)}
\label{Tabel 1}
\begin{tabular}{c c c c c c c c c c}
\hline
Model$^{\textcolor{red}{a}}$ & size & construction [s] & \multicolumn{3}{c}{Solution} & & \multicolumn{3}{c}{Total} \\
\cline{4-6} \cline{8-10}
  & (DoFs) & & NL [s] & APX [s] & speedup & & NL [s] & APX [s] & speedup   \\
 \hline
 RB & $52^{\textcolor{red}{b}}$ & 7.0761 & 2.5000 & 0.0016 & 1562.5 & & 9.5761 & 7.0777 & 1.3530\\ 
 SUB & $972^{\textcolor{red}{c}}$ & 0.2729 & 151.26 & 0.0899 & 1682.4 & & 151.53 & 0.3628 & 417.67\\
  \hline
  \multicolumn{10}{l}{$^{\textcolor{red}{a}}$ size of the full-order model: 5082}  \\
  \multicolumn{10}{l}{$^{\textcolor{red}{b}}$ 6 crack face forces (crack depth 15\%), 40 external forces, 6 modal coordinates}  \\
  \multicolumn{10}{l}{$^{\textcolor{red}{c}}$ substructure A: 926 (242 interface, 6 crack face forces, 678 internal DoFs)}\\
  \multicolumn{10}{l}{\hspace{1.5mm} substructure B: 294 (40 external forces, 242 interface, 6 modal coordinates)}
\end{tabular}
\end{table}

According to the table, the approximation method decreases the solution time by three orders of magnitude for both the RB and SUB models. The SUB model offers an advantage over the RB model, mainly due to differences in model construction time. Despite the SUB model's larger size compared to the RB model, its shorter online construction time results in a much faster overall online simulation than that of the RB model.

\subsubsection{Crack identification}

\begin{figure}[p]
    \centering
    \adjincludegraphics[width=5in,trim={0 0 {0.15\width} 0 },clip]{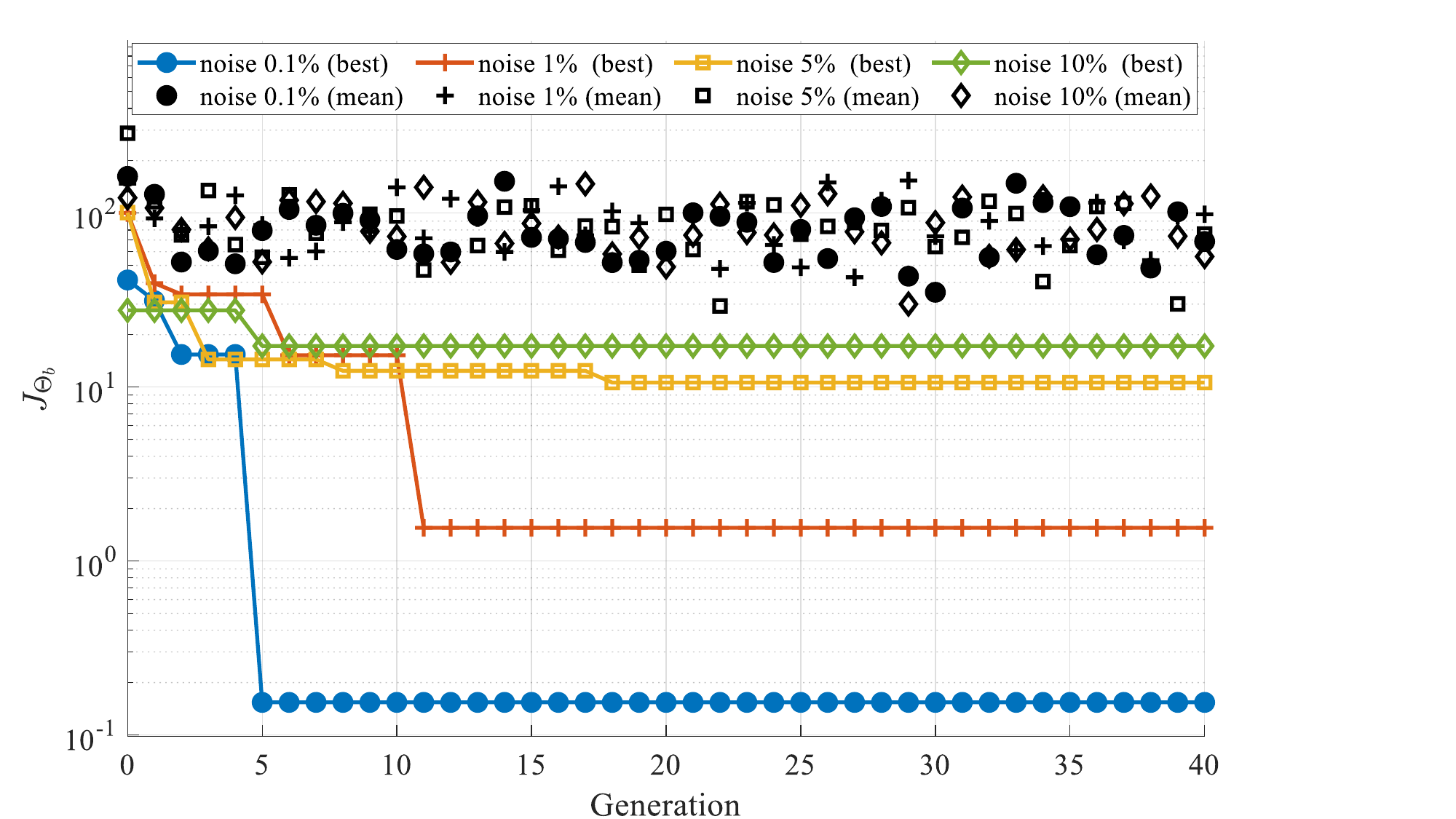}
    \caption{Best and mean fitness function ($J_{\Theta_{b}}$) value for each generation set of the GA algorithm ($\Theta_b$) for crack identification in 2D case under varying noise level of 0.1 to 10 present.}
    \label{fig: GA_2D}

    \centering
    \adjincludegraphics[width=6.5in,trim={0 0 {0.22\width} 0 },clip]{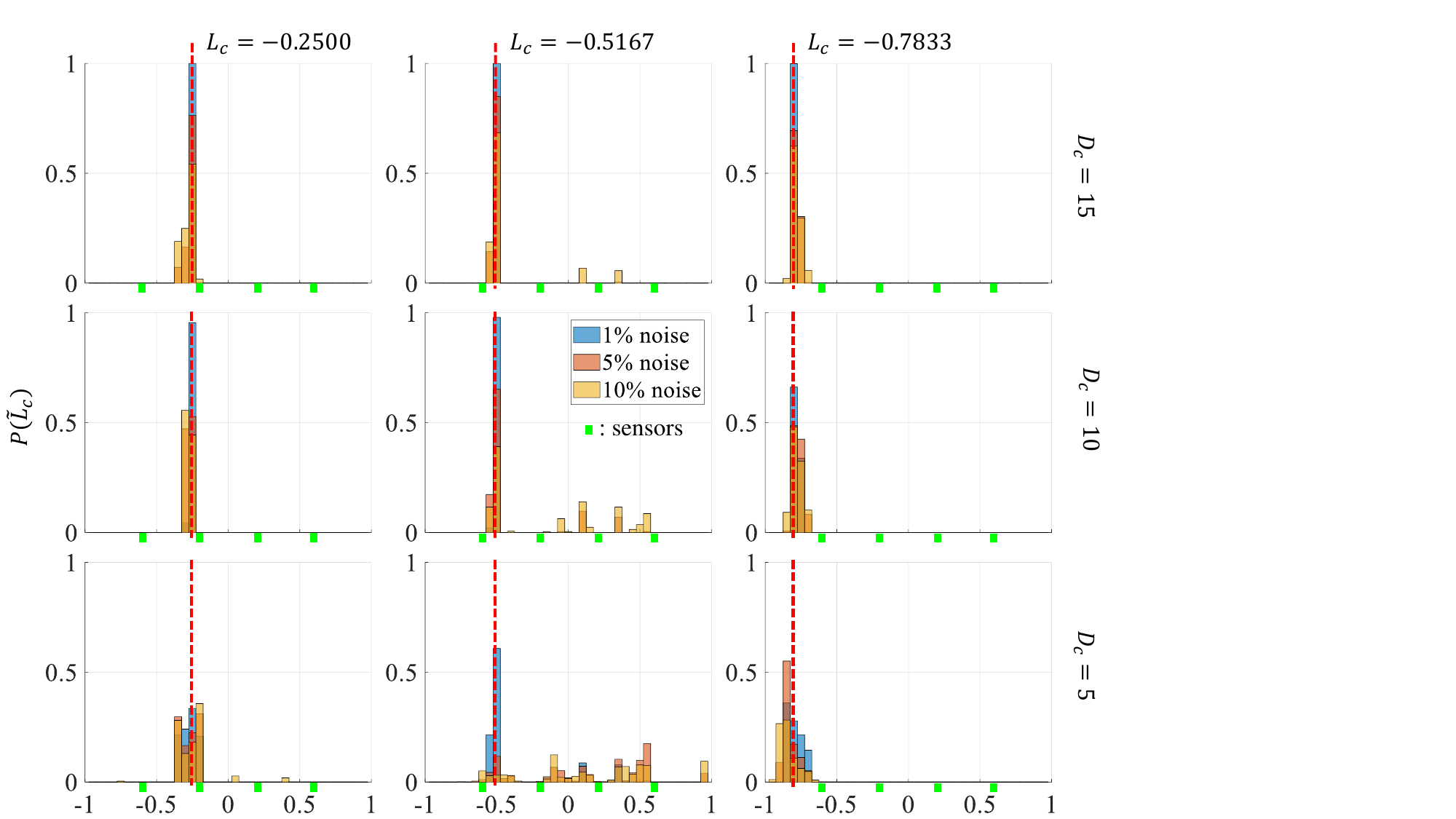}
    \caption{Probability distribution of the identified crack parameters for three different crack locations with different crack depth and noise level where $P(\tilde{L}_c)$ denotes the probability of identified crack location ($\tilde{L}_c$). True value of the crack ($L_c$) is marked with a dashed line.}
    \label{fig: statistic_2D}
\end{figure}

In this section, the approximated $\hat{\text{Tr}}^{(2)}$ is implemented for crack identification, following the procedure outlined in Section \ref{section: 2.5}. The loading and boundary conditions remain identical to those in the previous section. On the top region of the beam, four evenly spaced elements, identical to those mentioned in the previous section, function as virtual strain sensors, as depicted in Figure \ref{fig:TR_2D}.

Initially, a crack with $L_c = 0.5$, and $D_c = 10$ was examined, and the GA algorithm-based identification procedure was executed under different noise conditions. The adopted population size of the GA method is set to 7 individuals, and the total number of generations was capped at 40. Two members of each generation were guaranteed to remain in the new generation set through elitism. The results are presented in Figure \ref{fig: GA_2D}. As shown, when the noise level is 0.1\%, the GA method successfully converges to the true (reference) solution in fewer than 15 generations. As expected, higher noise levels hinder fast convergence.

To further assess the identification performance, a Monte Carlo analysis was conducted by executing the crack identification process 1000 times. This procedure was repeated for three different crack positions and three different crack depths, namely $D_c = \{5,10,15\}$ and $L_c = \{-0.7833,-0.5167,-0.2500\}$. The identified crack locations resulting from this study and their associated probabilities are plotted in Figure \ref{fig: statistic_2D}. As the figure illustrates, there is a clear positive correlation between the probability of correctly localizing the crack and the crack depth. For example, with $D_c = 15$, and 1\% noise, the true location of the crack is identified for all runs. The addition of noise was observed to deteriorate the accuracy of crack identification, especially in cases involving shallower cracks, highlighting the difficulty of detecting smaller cracks. Specifically, when \(L_c = 0.5\), under the extreme conditions of high noise (exceeding 5\%) and the smallest crack depth ($D_c = 5$), the method fails to accurately localize the crack. In the other two scenarios, the identified crack remains in close proximity to the actual location, even when the crack is small and the noise level is high.

\subsection{3D example}

\begin{figure}[h]
    \centering
    \adjincludegraphics[width=6in,trim={0 {0.4\height} {0.18\width} 0 },clip]{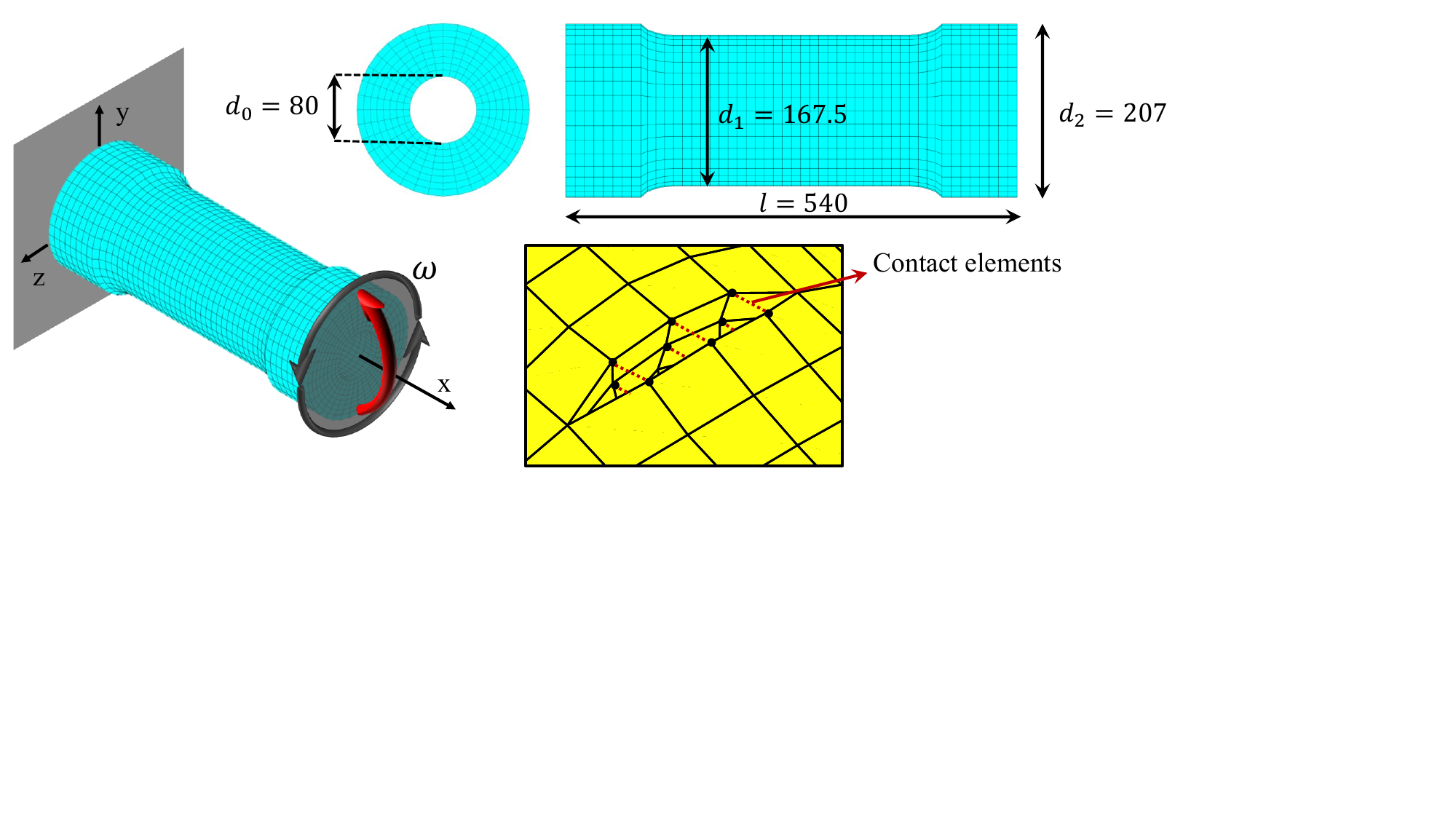}
    \caption{3D axle example. Dimensions are all in millimeter. Contact pairs at crack face are connected with dashed red lines.}
    \label{fig: 3D_model}
\end{figure}

For the 3D case, as shown in Figure \ref{fig: 3D_model}, a hollow axle with a varying cross-section, resembling a realistic train wheelset axle, is considered. The same material properties as in the 2D case are adopted. The axle is cantilevered at one end, and a rotating moment is applied at the other end. Similar to the 2D example, Rayleigh damping with a damping ratio of 0.002 was considered. The first three natural frequencies of the model are: $\omega_1$ = 483.20 Hz (1st bending), $\omega_2$ = 1490.55 Hz (1st torsional), and $\omega_3$ = 2162.89 Hz (2nd bending).

In the 3D case, the crack is parameterized with three spatial variables, namely, the normalized longitudinal location $L_c \in (0,1)$, circumferential location $\phi_c \in (-\pi,\pi)$, and the crack depth to the axle diameter ratio in percent $D_c \in \{3.83,7.67,11.50\}$. Because of FE discretization, the resulting parameter space was also discrete, while the size of this space became $\Theta \in \mathbb{R}^{49\times36\times3}$. The contact at the crack face is modeled by node-wise frictionless contact elements (see Fig. \ref{fig: 3D_model}) as discussed in section \ref{section: Methods}. 

\begin{figure}[h]
    \centering
    \adjincludegraphics[width=6in,trim={0 {0.54\height} {0.05\width} 0 },clip]{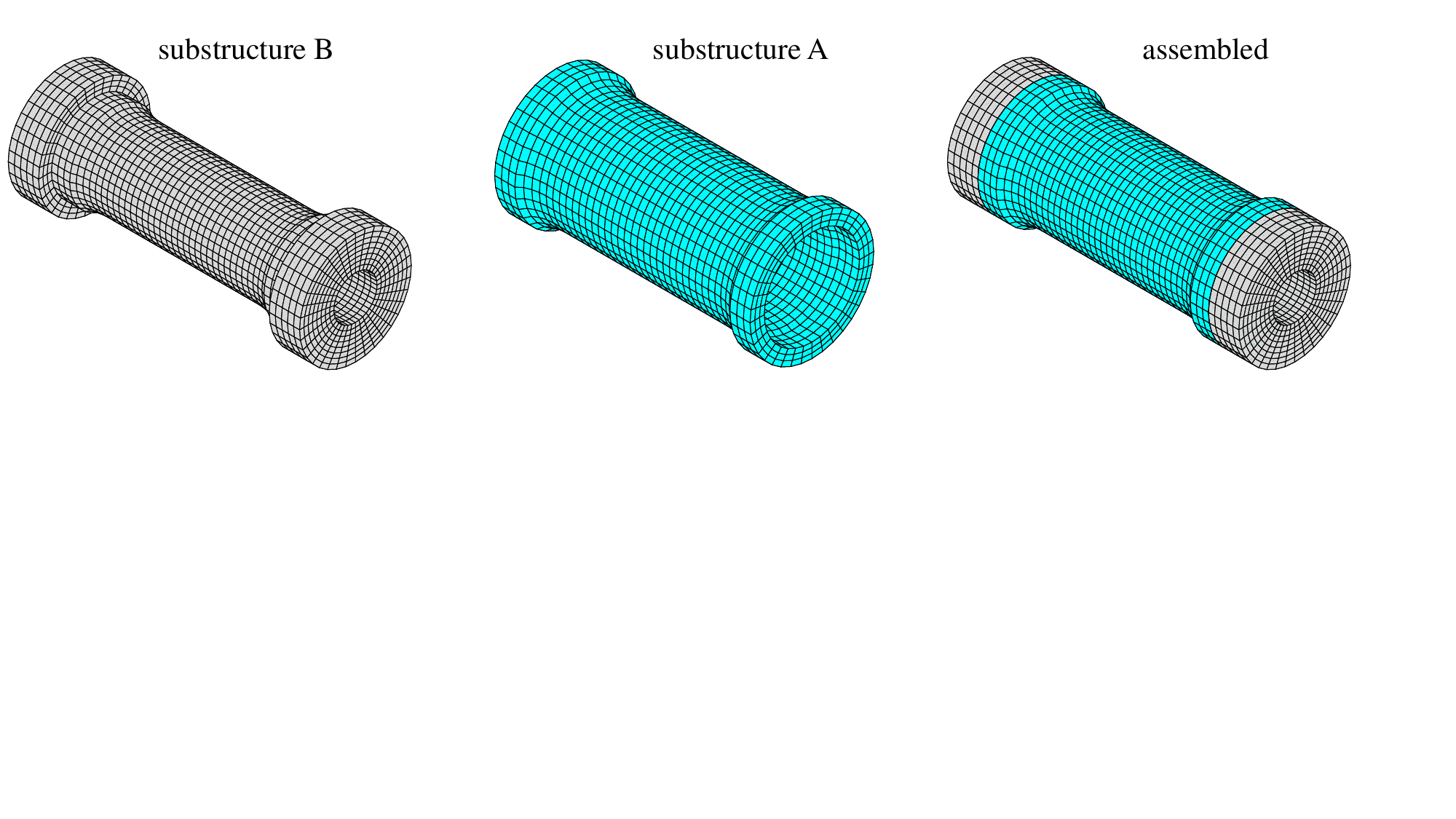}
    \caption{Components of the SUB model for the 3D example.}
    \label{fig:3model - 3D}
\end{figure}

In line with the previous example, we present two reduced-order models aimed at speeding up simulations for the 3D case. The RB model was derived by keeping the crack face DoFs and external forcing DoFs, and reducing all other DoFs using the Rubin reduction technique. For the SUB model, the structure is divided into two substructures. Substructure A is a subpart of the structure where cracks are assumed to occur. In this case, substructure A becomes the skin of the axle, as shown in Figure \ref{fig:3model - 3D}, while Substructure B contains the remaining part of the domain. As done previously, substructure B is reduced to the interface DoFs and external forcing DoFs via the Rubin method.

 \subsubsection{Sensitivity of HOTr to the crack parameters}
 
\begin{figure}[h]
    \centering
    \adjincludegraphics[width=7in,trim={0 {0.3\height} 0 0 },clip]{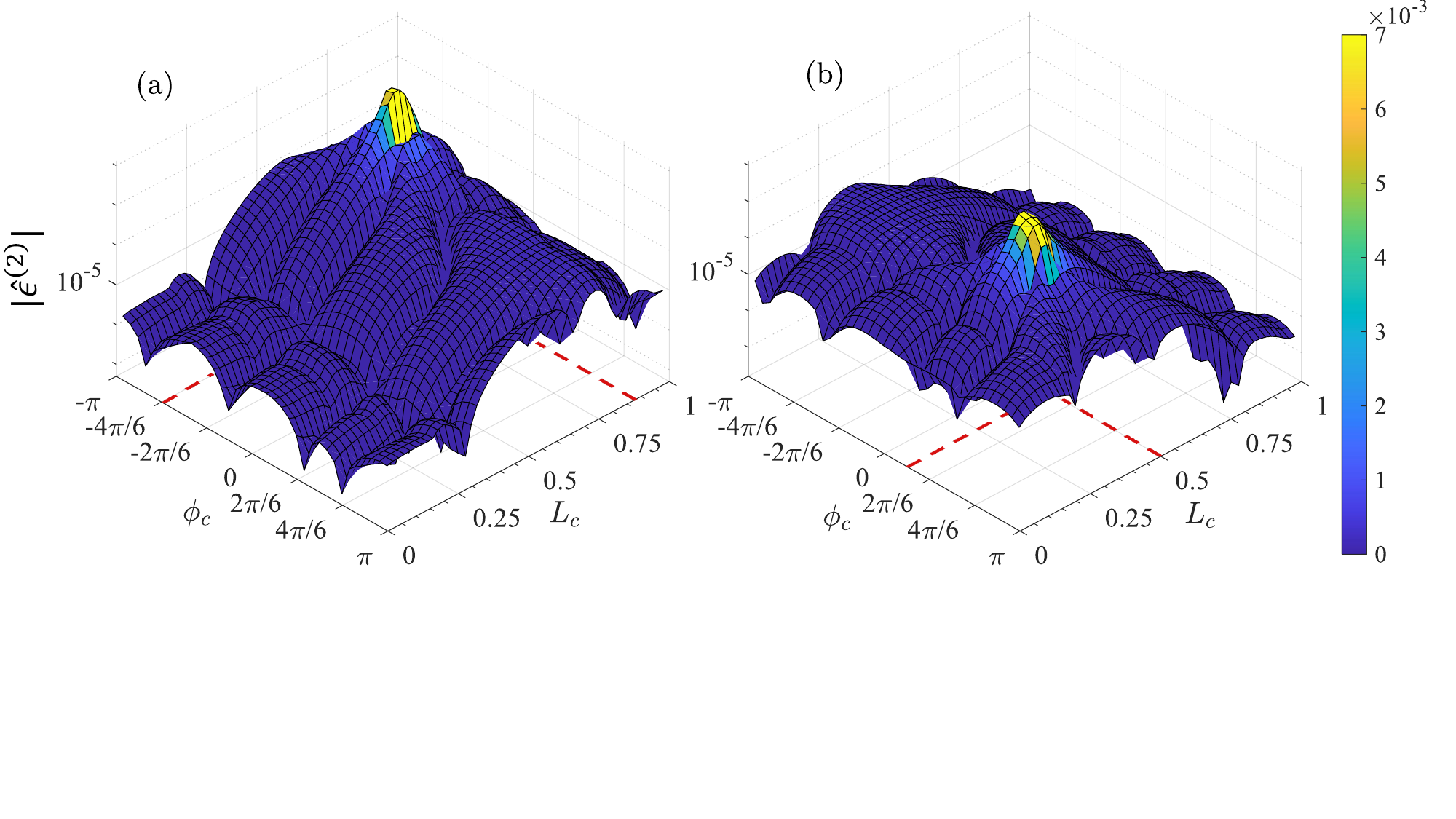}
    \caption{Second-order harmonic strain $\hat{\epsilon}^{(2)}$ on the surface of the beam in the vicinity of the crack. (a): Crack parameters: $L_c = 0.875$, $\phi = -4\pi/6$. (b): Crack parameters: $L_c = 0.5$, $\phi = \pi/6$.}
    \label{fig: FRF_3D}
\end{figure}

To assess the sensitivity of the second-order harmonic feature to the crack parameters, the second-order harmonic strains $\hat{\epsilon}^{(2)}$, in the $x$ direction on the axle surface, are plotted in Figure \ref{fig: FRF_3D}. The procedure for deriving the steady-state response using the MHB method is analogous to the 2D case and is carried out on the RB reduced-order model. 

Figure \ref{fig: FRF_3D} clearly demonstrates the high sensitivity of $\hat{\epsilon}^{(2)}$ to the location of the crack. Variations in the crack position induce notable changes in the $\hat{\epsilon}^{(2)}$ profile. As expected, the magnitude of $\hat{\epsilon}^{(2)}$ rises close to the crack. Analogous to the 2D example, the anti-peaks of $\hat{\epsilon}^{(2)}$ (which transform into valleys shown in Figure \ref{fig: FRF_3D}) adjust as the crack location shifts.

It is worth reminding that higher-order harmonics of the response, such as $\hat{\epsilon}^{(2)}$, are a consequence of the breathing crack phenomenon. In a pristine axle, these higher-order strain harmonics would be absent. Consequently, as illustrated in Figure \ref{fig: FRF_3D}, a nonzero second-response reading — indicating the presence of a crack — can be detected even from a distant sensor measurement. However, accurately pinpointing the location of the crack requires additional information, which will be discussed in detail in section \ref{section: 3.2.3}.

\subsubsection{Approximation of the HOTrs}
\label{section: 3.2.2}

\begin{figure}[h]
    \centering
    \adjincludegraphics[width=7in,trim={0 0 {0.075\width} {0.02\height} },clip]{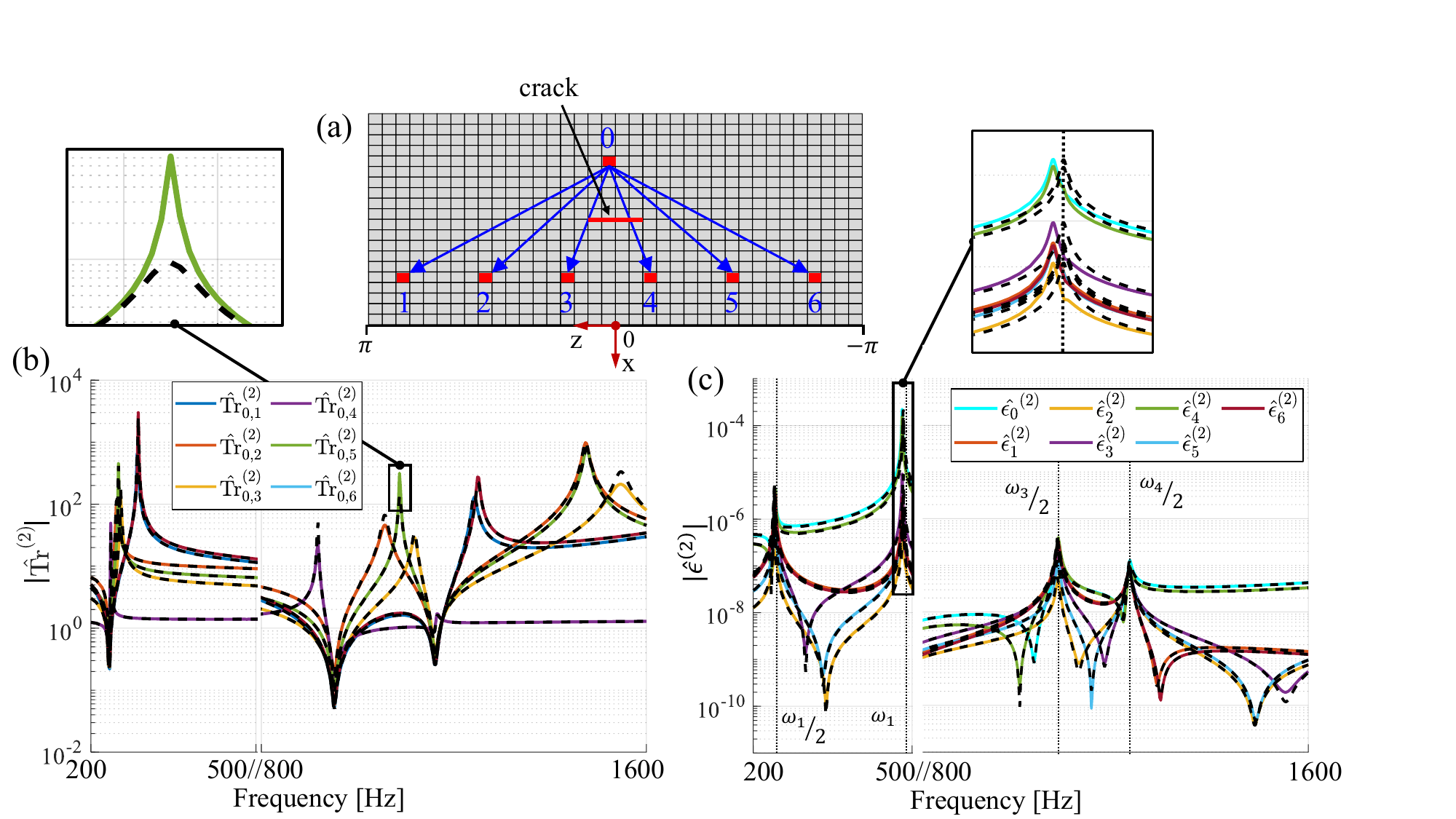}
    \caption{Second-order transmissibility ($\hat{\text{Tr}}^{(2)}$) and corresponding second-order strains ($\hat{\epsilon}^{(2)}$); solid lines represent nonlinear solution obtained with MHB, dashed lines the approximation using the proposed method. (a) illustrates the flatten surface around crack. points at which the HOTr is calculated and its corresponding directions. (b) $\hat{\text{Tr}}^{(2)}$ in the direction of point 0 to points 1 to 6 shown with blue color arrows in (a). Maximum and average RMSE for all six curves in this figure are 0.2214, and 0.0054 respectively. (c) $\hat{\epsilon}^{(2)}$ in the direction of point 5 to points 7 to 12 shown with back colored arrows in (a). Maximum and average RMSE for all six curves in this figure are 0.2555, and 0.017 respectively. }
    \label{fig: TR_3D}
\end{figure}

To validate the proposed approximation method for deriving the HOTr, the second-order transmissibility of \emph{strain} readings (denoted as $\hat{\text{Tr}}^{(2)}$) is computed. The strains are measured in the $x$ direction. The locations of the measurement points and the directions of the transmissibility functions are shown in Figure \ref{fig: TR_3D}(a). As shown in Figure \ref{fig: TR_3D}(b) and Figure \ref{fig: TR_3D}(c), the proposed method approximates $\hat{\text{Tr}}^{(2)}$, along with the second-order harmonic of strains ($\hat{\epsilon}^{(2)}$), with very good accuracy, as only minor discrepancies are noted in the zoomed sub-figures. Expectedly, the resonance frequency is slightly overestimated in the approximate result (Figure \ref{fig: TR_3D}(c)).

As mentioned earlier, the external force exerted on the axle is assumed to be due to the rotation of the axle under a constant load of the train weight, creating harmonic forcing with a frequency proportional to the speed of the train. For a typical design, the resulting frequency is lower than the first natural frequency of the axle. However, the precision of the proposed method is tested on a larger frequency band that spans up to 1600 Hz. Thus, considering the resonances of the first two vibration modes.

Analogous to the two-dimensional scenario, Table \ref{table: 2} details the duration of both constructing and solving the model during a single-forward evaluation. In this particular test, $D_c = 11.5$ was considered with an excitation frequency of 100 Hz. The simulations are carried out on the same local computer as for the 2D case.

\begin{table}
\caption{Computational time comparison between the nonlinear solution (NL), and proposed approximation (APX)}
\label{table: 2}
\begin{tabular}{c c c c c c c c c c}
\hline
$\text{Model}^{\textcolor{red}{a}}$ & size &  Construction [s] & \multicolumn{3}{c}{Solution} & & \multicolumn{3}{c}{Total} \\
\cline{4-6} \cline{8-10}
  & (DoFs) & & NL [s] & APX [s] & speedup & & NL [s] & APX [s] & speedup   \\
 \hline
 RB & $350^{\textcolor{red}{b}}$ & 33.685 & 78.438 & 0.0240 & 3268.2 & & 112.12 & 33.710 & 3.3260\\ 
 SUB & $18048^{\textcolor{red}{c}}$ & 3.0776 & - & 4.1088 & - & & - & 7.1864 & -\\
  \hline
  \multicolumn{10}{l}{$^{\textcolor{red}{a}}$ size of full-order model: 49593 DoFs}  \\
  \multicolumn{10}{l}{$^{\textcolor{red}{b}}$ 14 crack face forces (crack depth: 11.5\%), 324 external forces, 12 modal coordinates}  \\
  \multicolumn{10}{l}{$^{\textcolor{red}{c}}$ substructure A: 17712 (5076 interface, 14 crack face forces, 12622 internal DoFs)}\\
  \multicolumn{10}{l}{\hspace{1.5mm} substructure B: 5412 (324 external forces, 5076 interface, 12 modal)}
 
\end{tabular}
\end{table}

Similar to the 2D example, the suggested method significantly accelerates the solution process for the 3D example. As shown in Table \ref{table: 2}, the \emph{solution} time for the RB model is reduced by a factor of 3268.2. It should also be mentioned that, due to the large size of the SUB model, its nonlinear solution through the MHB method is practically infeasible, while it takes only around 7 seconds for the proposed method to calculate its approximation. Given the increased size of the model in the 3D example compared to that in the 2D example, the model construction time plays a more prominent role in the overall computational cost, further highlighting the effectiveness of the SUB model over the RB model. As shown, the \emph{total} solution time required for the SUB model is approximately $4.7 = {}^{33.710}{\mskip -5mu/\mskip -3mu}_{7.1864}$ times less than that for the RB model, despite the fact that the size of the RB model is smaller.

\subsubsection{Crack identification}
\label{section: 3.2.3}

Finally, we considered the 3D axle model for crack identification. The excitation frequency was set to 100 Hz. It should be mentioned that, since the accuracy of the proposed method is guaranteed for the higher frequencies, as shown in the previous section, there is no limitation on applying the method in the higher frequency range. The reason for keeping the frequency in a lower range is to resemble a real-world scenario where the angular velocity of the axle is lower than its first natural frequency. 

In this case, the size of the GA population was set at 10, and the maximum number of generations was capped at 50. The size of the elite population was kept at two.

\begin{figure}[h]
    \centering
    \adjincludegraphics[width=6in,trim={0 {0.45\height} {0.24\width} 0 },clip]{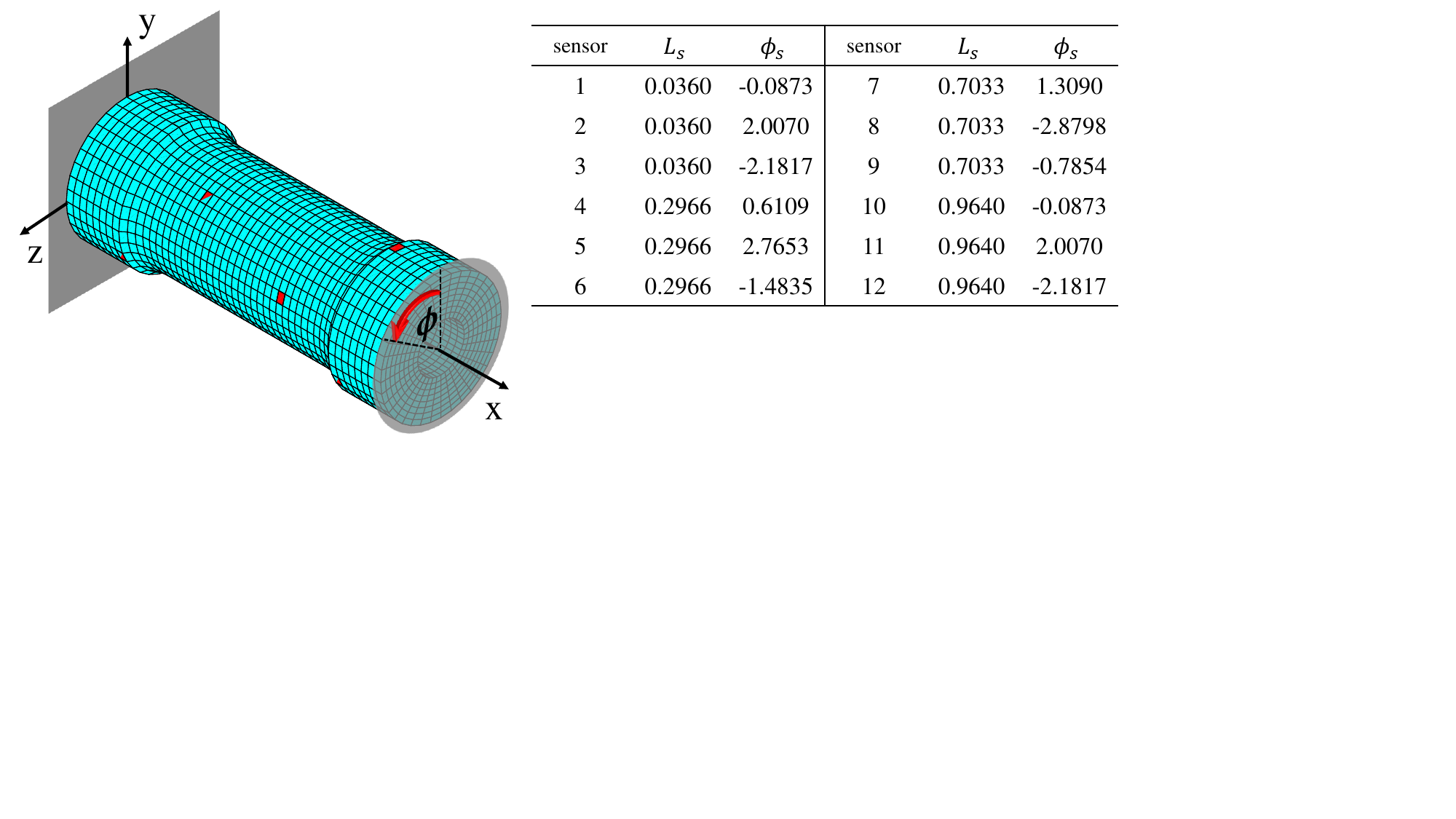}
    \caption{Virtual strain sensor locations for the 3D model.}
    \label{fig: sensors_3D}
\end{figure}

For this analysis, 12 sensors were chosen at 4 different cross sections uniformly distributed across the length of the axle. The three sensors in each cross section are positioned at equal circumferential distances from each other. However, their orientations differ from one another, as shown in \ref{fig: sensors_3D}. The reason for this is to minimize the blind spot of the sensors in the circumferential direction. The strain reading in the $x$ direction of the axle is used as the measured signal. Finally, for the objective function (Eq. \ref{Eq. objective function}), $\hat{\text{Tr}}^{(2)}$ is calculated between all virtual sensors and in both directions.

\begin{figure}[H]
    \centering
    \adjincludegraphics[width=4.5in,trim={0 0 {0.18\width} 0 },clip]{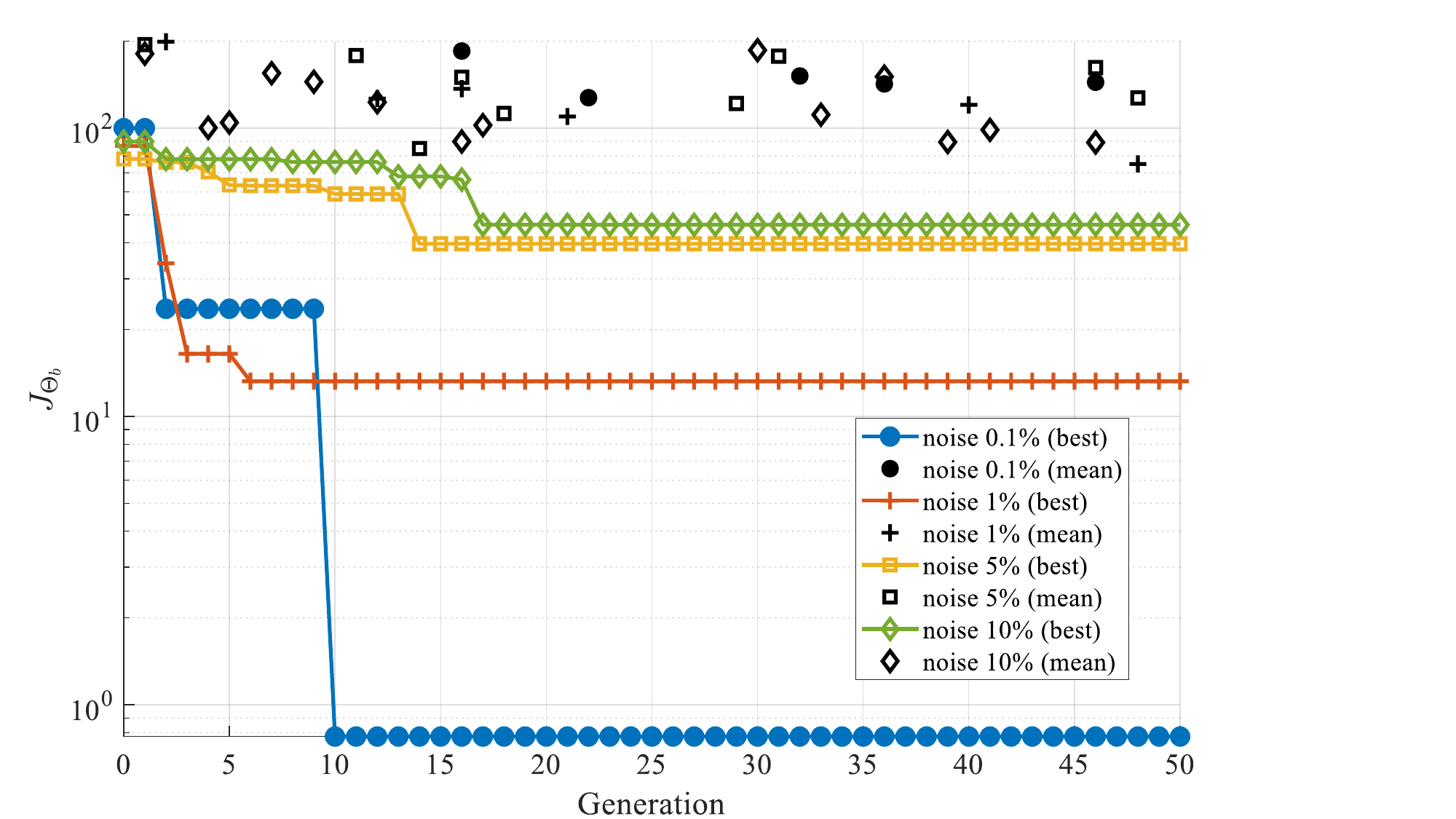}
    \caption{Best and mean fitness function ($J_{\Theta_{b}}$) value for each generation of the GA algorithm for crack identification in 3D case under varying noise level. Sub-figures a-d represent noise levels of 0,1,5,10 percent respectively.}
    \label{fig: GA_3D}
\end{figure}

Similarly to the 2D case, for the first step, we evaluated the GA algorithm by performing crack identification for a specific crack location and different noise levels. The target crack position was $L_c = 0.3102, \phi_c = 1.0472, D_c = 7.67\%$. A similar trend is observed as in the 2D case. For lower noise levels, identifying the crack location is more straightforward than in the case of high noise contamination. This can be understood from the sudden drop in the penalty value of the case with a higher signal-to-noise ratio in Figure \ref{fig: GA_3D}. However, we observed that the GA method quickly converges to a parameter set in a region close to the original crack parameters, and due to high noise contamination, later generations struggle to improve the location of the identified crack.

\begin{figure}[h]
    \centering
    \adjincludegraphics[width=6.4in,trim={0 {0.11\height} 0 0 },clip]{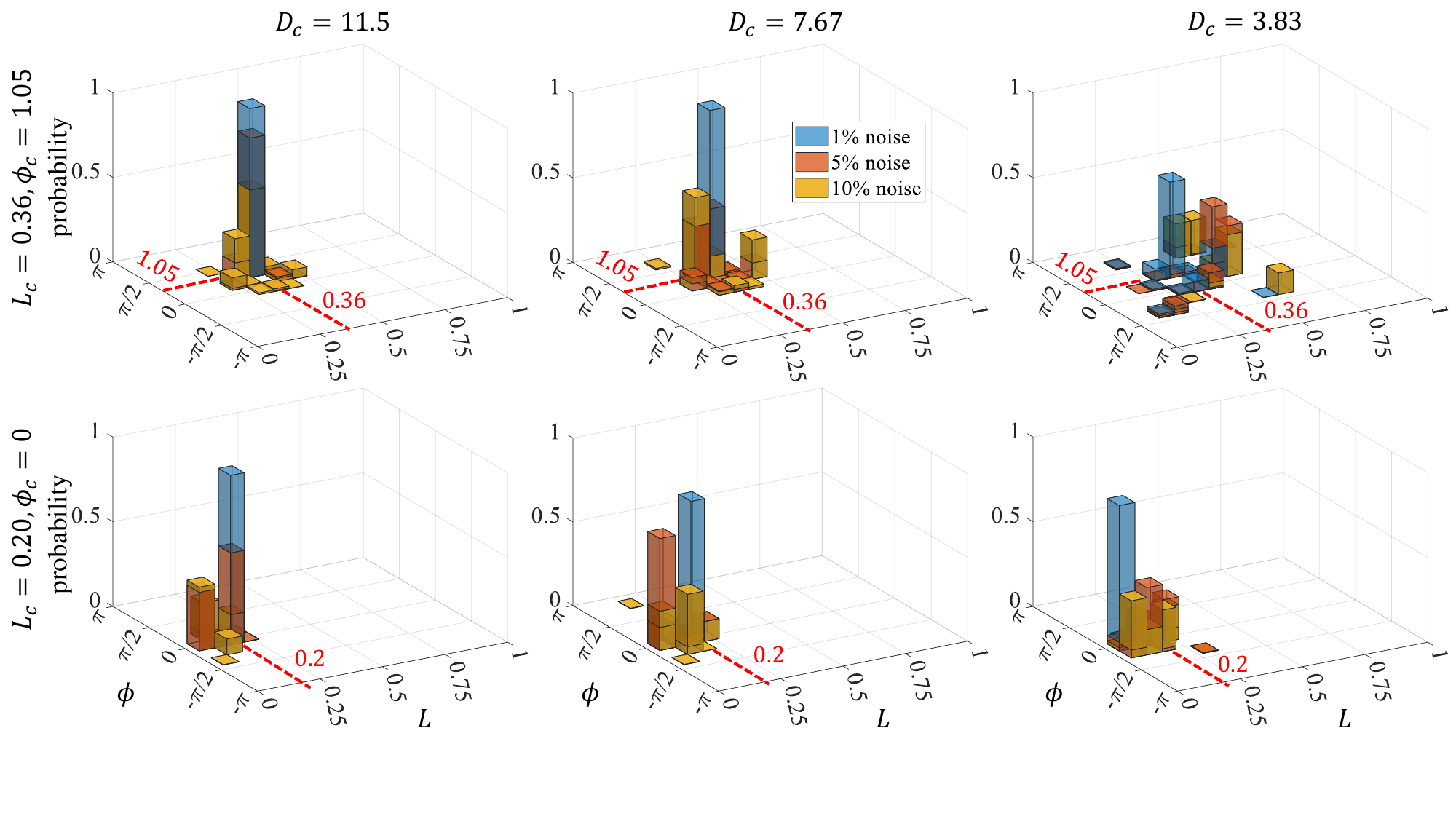}
    \caption{Statistical view on the identified crack parameters for two different crack locations with different crack depth and noise level}
    \label{fig: statistic_3D}
\end{figure}

Next, Monte Carlo simulations are conducted to assess the performance of the identification process. For this, two different crack scenarios are examined. For each case, three different crack depths are considered, and three different noise levels are added to the simulated measurements. Noting that in each case, crack depth was imposed, and its location \{$L_c, \phi_c$\} was identified. Figure \ref{fig: statistic_3D} illustrates the results of the Monte Carlo simulation, which involved 1000 iterations for each identification case. The balance between the crack depth and the noise level for accurate crack identification was also observed in the 3D case. As mentioned above, despite the lower probability of exact parameter identification, in almost all cases, the identified parameters lie in close proximity to the target.

\section{Conclusions}
\label{section: conclusion}

In this study, we consider the higher-order harmonics arising from the breathing of cracks as a potential feature for crack identification. We observe that for breathing crack phenomena, the second-order harmonic delivers a prominent damage-sensitive feature. Thus, we propose second-order transmissibility $\hat{\text{Tr}}^{(2)}$ as a crack identification feature for in-service health monitoring. In line with previous reports in the literature (\cite{sinou2009review,lin2018higher}), it is observed that, in contrast to their first-order counterparts, second-order harmonics prove to be highly sensitive to variations in crack location and size (see Figure \ref{fig: FRF_2D}). Moreover, these features can be detected from sparse measurements, making them efficient crack identification features.

One major bottleneck in implementing nonlinear features for crack identification is the associated high computational cost. This is primarily because the identification procedure requires numerous forward solutions of the computational model, which, given the nonlinear nature of the higher-order harmonics, necessitates an iterative solution procedure (e.g. Newton-Raphson). To alleviate this toll, we propose an approximation for HOTr that eliminates the need for the expensive nonlinear solution process. The main idea of the proposed method is to approximate the nonlinear crack face forces with the linear solution to the constraint forces that keep the crack closed. This method works well for incipient cracks where the overall response of the cracked axle is not significantly different from that of the pristine one. Therefore, the method is not expected to work for large cracks. The proposed method reduces the \emph{solution} time by three orders of magnitude for both 2D and 3D cases. The accuracy of the proposed method is tested for 2D and 3D cases, where the ratio of crack depth to axle diameter is 15\% and 11.5\%, respectively. For the 2D case, the method achieves excellent accuracy across the entire frequency range. The accuracy for the 3D case is reduced; however, overall, the method achieves good accuracy; one that proves sufficient for the identification task.

Another computational challenge in calculating the HOTr is model construction. Since the nonlinear force is confined to crack faces, Reduced-Order Modeling (ROM) techniques, like the Rubin method, are very appealing. However, a new ROM must be constructed for every new set of crack parameters, making the procedure ineffective.  To overcome this challenge, we used a substructuring technique to reduce the partition of the model that is not affected by the crack once for all parameters and kept the other subdomain unreduced. The combination of the proposed approximation method for $\hat{\text{Tr}}^{(2)}$ and this substructuring technique significantly accelerates the derivation of the HOTr.

Finally, crack identification is performed on 2D and 3D train wheelset axle models. The efficiency of the proposed technique is studied through Monte-Carlo analysis. As expected, a trade-off is noted between crack size, noise level, and localization precision. Although the probability of correct localization in the case of cracks larger than 10\% and noise levels less than 5\% is almost 1, precision drops for smaller cracks and higher noise levels. Nevertheless, as shown in Figures \ref{fig: statistic_2D} and \ref{fig: statistic_3D}, the identified crack parameters lie in close vicinity of the target. 

The proposed procedure can be readily applied to any structure subjected to harmonic excitation, where a breathing crack represents the dominant source of nonlinearity, without requiring any modifications or additional considerations. However, it should be noted that the present study focuses on the case in which the breathing crack is the dominant source of nonlinearity. In situations where additional nonlinear effects are comparatively weak, the proposed methodology remains valid, as these effects can be treated as secondary perturbations or effectively as noise. Conversely, when other nonlinearities become significant (e.g., under large deformations), their influence must be explicitly incorporated into the computational model. In such cases, the framework can theoretically be expanded, though this would add to the model’s complexity and highlight the necessity of developing efficient reduced-order formulations for practical identification.

An acknowledged limitation of this study is its reliance on a mesh-based model for crack geometry. Future research could extend this work by employing meshless techniques, such as XFEM, for improved crack representation. Additionally, experimental validation of the provided results is crucial. The authoring team is currently conducting a dedicated experimental campaign to evaluate the proposed methodology under controlled laboratory conditions. The results of this ongoing work will be presented in a forthcoming publication focused specifically on the experimental verification of the HOTr-based crack identification framework.

\paragraph{Declaration of Competing Interest\\}
\addcontentsline{toc}{section}{Declaration of Competing Interest}
The authors declare that they have no known competing financial interests or personal relationships that could have appeared to influence the work reported in this paper.

\paragraph{Acknowledgments\\}
\addcontentsline{toc}{section}{Acknowledgements}
The authors acknowledge the funding of the ETH Center for Sustainable Future Mobility project “SENTINEL - In-SErvice diagnostics of the cateNary/panTograph and wheelset axle systems through INtELligent algorithms,” within the context of ETH Mobility grants.

\bibliographystyle{unsrt}
\bibliography{references}

\end{document}